\documentclass[11pt]{article}
\usepackage[margin=1in,headheight=14pt]{geometry}
\usepackage[T1]{fontenc}
\usepackage{lmodern,microtype,amsmath,amssymb,amsthm,mathtools,bm}
\usepackage{graphicx,booktabs,tabularx,array,enumitem,needspace,placeins,flafter}
\usepackage{tikz}
\usetikzlibrary{arrows.meta,positioning,calc}
\usepackage[numbers,sort&compress]{natbib}
\usepackage{xurl}
\usepackage[hidelinks]{hyperref}
\usepackage[nameinlink,noabbrev,capitalize]{cleveref}
\usepackage{fancyhdr,aliascnt}
\allowdisplaybreaks[1]
\newtheorem{theorem}{Theorem}
\newaliascnt{proposition}{theorem}
\newtheorem{proposition}[proposition]{Proposition}
\aliascntresetthe{proposition}
\crefname{proposition}{Proposition}{Propositions}
\Crefname{proposition}{Proposition}{Propositions}
\newaliascnt{corollary}{theorem}

\aliascntresetthe{corollary}
\crefname{corollary}{Corollary}{Corollaries}
\Crefname{corollary}{Corollary}{Corollaries}
\newaliascnt{lemma}{theorem}
\newtheorem{lemma}[lemma]{Lemma}
\aliascntresetthe{lemma}
\crefname{lemma}{Lemma}{Lemmas}
\Crefname{lemma}{Lemma}{Lemmas}
\newtheorem*{corollary*}{Corollary}
\newtheorem{suppproposition}{Supplementary Proposition}

\theoremstyle{definition}
\newaliascnt{definition}{theorem}
\newtheorem{definition}[definition]{Definition}
\aliascntresetthe{definition}
\crefname{definition}{Definition}{Definitions}
\Crefname{definition}{Definition}{Definitions}
\theoremstyle{remark}
\newaliascnt{remark}{theorem}

\aliascntresetthe{remark}
\crefname{remark}{Remark}{Remarks}
\Crefname{remark}{Remark}{Remarks}
\newcommand{\E}{\mathbb E}
\newcommand{\Prob}{\mathbb P}
\newcommand{\R}{\mathbb R}
\newcommand{\N}{\mathbb N}
\newcommand{\ind}{\mathbf 1}
\newcommand{\TT}{\mathcal T}

\newcommand{\HH}{\mathcal H}
\newcommand{\PP}{\mathcal P}
\newcommand{\RR}{\mathsf R}
\newcommand{\norm}[1]{\left\lVert#1\right\rVert}

\newcommand{\KL}{D_{\mathrm{KL}}}
\newcommand{\TV}{\operatorname{TV}}
\newcommand{\TC}{\operatorname{TC}}
\newcommand{\tr}{\operatorname{tr}}

\newcommand{\Ber}{\operatorname{Ber}}
\newcommand{\Unif}{\operatorname{Unif}}
\newcommand{\dist}{\operatorname{dist}}
\newcommand{\diag}{\operatorname{diag}}

\newcommand{\conv}{\operatorname{conv}}
\newcommand{\op}{\mathrm{op}}
\newcommand{\negl}{\operatorname{negl}}

\hypersetup{pdftitle={Why Finite Intelligence Works in an Infinite World: A Mathematical Theory},pdfauthor={Mojtaba Soltanalian}}

\begin{document}

\title{\vspace{-1.5em}\textbf{Why Finite Intelligence Works in an Infinite World:}\\[.3em]\textbf{A Mathematical Theory}}
\author{Mojtaba Soltanalian\\\small University of Illinois Chicago}
\date{November 13, 2026}
\maketitle
\begin{abstract}
An information system often has to learn before it knows what it will be asked.  A finite representation can succeed without recording an infinite environment in full when those asks share collective structure, rather than demand unrelated facts scattered arbitrarily across it. We formalize this phenomenon through future-task information, the minimum retained information needed to preserve decision value relative to the available evidence. A joint decision-regret state records posterior comparisons among actions for every permitted task. Under compactness assumptions, every retained representation admits a calibrated replacement based on this state, preserving all task regrets without increasing information; the resulting information--regret curve equals an operational coding rate. Two exact boundaries govern finite preservation: weak-star closure for a fixed action count, and the singular mass of the source--evidence law when all finite action menus are admitted. Collective structure determines information requirements beyond individual-question difficulty. Families with identical source laws, pairwise task metrics, and individual-task rates can have different high-resolution exponents. Covariance-identical coordinate and parity families have different exact rate--distortion curves, with optimal parity retention characterized by a sharp Fourier--entropy inequality. For balanced nested task families, refinement sizes and weighted reach determine the rate within an additive constant, attained by prefix coding under explicit decay conditions. Nonlinear learning and computational separations distinguish preservation from efficient discovery and use. The proposed mathematical theory identifies which shared decision requirements allow finite information to remain useful for infinitely many possible purposes.
\end{abstract}

\section{Learning before the question}
\label{sec:intro}
Information is usually acquired for uses that are only partly known.  A recommendation system learns from past interactions before the next request arrives. A language model is trained before it sees the next question. And work on one computational model may precede the specification of another. In each case, preparation precedes purpose.

The ordering changes the compression problem. When the future decision is known, preparation may retain only its preferred action. When the decision is announced later, the same retained state must support different actions under different losses. A finite system succeeds when many possible uses depend on a common set of distinctions. It fails when the future family can repeatedly ask for distinctions that its retained state has discarded.

The application areas in \cref{tab:motivation} respond to preferences, questions, or models not yet specified. Their shared feature is the order of preparation and use; their observation models and algorithms may differ.

\begin{table}[htbp]
\centering
\small
\renewcommand{\arraystretch}{1.22}
\begin{tabularx}{\linewidth}{@{}>{\raggedright\arraybackslash}p{.29\linewidth}>{\raggedright\arraybackslash}X>{\raggedright\arraybackslash}X@{}}
\toprule
\textbf{Setting} & \textbf{Prepared in advance} & \textbf{Specified later}\\
\midrule
\textbf{Recommendation}\newline Netflix / MovieLens~\cite{koren2009,harper2015}
 & User--item interactions, observed preferences, and learned relations among users and items
 & The requested recommendation, relevant context, ranking objective, and cost of an incorrect prediction\\
\addlinespace[3pt]
\textbf{Large language models}~\cite{vaswani2017,brown2020}
 & Parameters and reusable representations learned from broad training data
 & The question or prompt, desired response, constraints, and criterion for an acceptable answer\\
\addlinespace[3pt]
\textbf{AI model construction}\newline Weight-space programming~\cite{takehi2026}
 & Previously learned weights, parameter relationships, model structure, and reusable computational results
 & The model to be built, its architecture and objective, training constraints, and required quality or guarantees\\
\bottomrule
\end{tabularx}
\caption{Three forms of preparation before purpose. Recommendation, language-model use, and model construction all retain information for later requests. Each setting has its own observation model, future-task family, loss, and computational demands.}
\label{tab:motivation}
\end{table}
\FloatBarrier

The issue depends on which distinctions future requests share. A finite procedure can evaluate infinitely many inputs. Conversely, if independent fair bits are assigned to infinitely many requests, every particular bit may be observable while no finite common state serves all requests uniformly. The question is therefore which information must survive, relative to a specified observation experiment and a specified family of possible uses.

Here \emph{finite intelligence} means finite retained information that preserves decision performance for tasks announced after preparation. The theory measures that performance by excess Bayes loss relative to the available full evidence. This benchmark matters. An uncertain world and a damaged representation are different sources of error: full evidence may itself leave uncertainty, and compression is charged only for the additional loss it creates.

\subsection{The central distinction}
A natural first approach assigns a distance to two worlds by the largest difference any future task can detect. That distance governs retention in regular models; more general families also depend on how the tasks are arranged together. Two task families can separate exactly the same pairs of worlds and present equally difficult individual questions, yet require different amounts of common retained information. The missing ingredient is how the questions are arranged jointly across the source distribution.

We retain this joint structure. For each observation, we record the excess posterior loss of every action in every permitted task. The resulting \emph{decision-regret state} removes task-wise offsets that cannot affect a choice, while keeping the comparisons on which future decisions depend. Any retained representation can be replaced, without increasing information or regret, by a calibrated compression of this state. The information required to attain uniform regret at most $D$ is denoted~$\RR_{\TT}(D)$.

The main argument follows three questions. What information must survive when its future purpose is unknown? Why do pairwise geometry and second-order statistics fail to determine its cost? Under what structural conditions can that cost be calculated and attained? The boundary theorems answer when finite preservation is possible; the acquisition and computation results then separate preservation from efficient discovery.

\begin{figure}[t]
\centering
\begin{tikzpicture}[>=Latex, node distance=7mm, every node/.style={font=\small}, box/.style={draw,rounded corners=1pt,align=center,minimum height=13mm,text width=22mm}]
\node[box] (world) {World\\$\Theta$};
\node[box,right=of world] (evidence) {Evidence\\$Y$};
\node[box,dashed,right=of evidence] (state) {Decision state\\$X$};
\node[box,right=of state] (memory) {Retained state\\$U$};
\node[box,right=of memory] (action) {Response\\$a_t(U)$};
\draw[->] (world)--(evidence);
\draw[->] (evidence)--(state);
\draw[->] (state)--(memory);
\draw[->] (memory)--(action);
\node[above=9mm of action,align=center] (task) {Future task $t$};
\draw[->] (task)--(action);
\draw[densely dashed] ($(memory.north east)+(3mm,5mm)$)--($(memory.south east)+(3mm,-4mm)$);
\node[below=6mm of state,align=center] {Preparation: the task is unknown};
\node[below=6mm of action,align=center] {Use: the task is given};
\end{tikzpicture}
\caption{The ordering studied throughout the paper. Evidence is encoded before the future task is known. The dashed decision-state box represents a mathematical reduction of the evidence; its computational realization depends on the observation model and task family. Every source-dependent quantity available at deployment is charged to the retained state.}
\label{fig:ordering}
\end{figure}
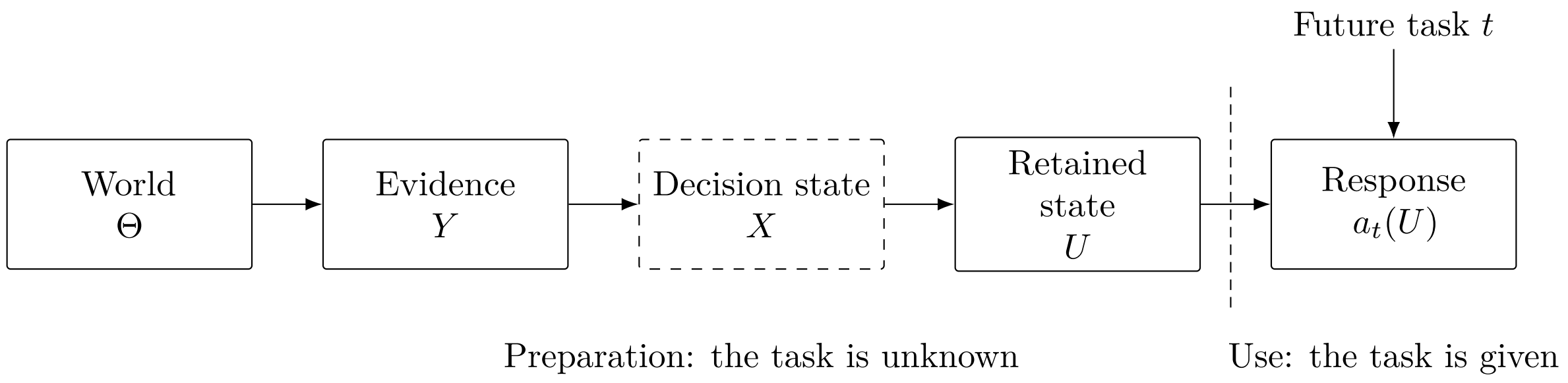

\subsection{Relation to information and decision theory}
Shannon's theory makes the communication and compression of information an operational question~\cite{shannon1948,shannon1959}. Blackwell compares experiments by the decisions they support \cite{blackwell}; Le Cam's theory studies approximation in terms of decision performance \cite{lecam}. Kanaya and Nakagawa connect mutual information with Bayes risk \cite{kanaya}. Indirect source coding distinguishes the observations available to an encoder from the quantity relevant to its decoder \cite{wolfziv}. Information bottleneck and deficiency-based feature learning develop corresponding notions of relevance and loss preservation \cite{tishby,rooyen2014,rooyen2015}. Posterior splitting and Bayes plausibility also have a central role in information design \cite{kamenica}. Rate--distortion methods likewise quantify minimum excess risk in Bayesian learning \cite{hafez}.

Several close antecedents isolate parts of the present problem. Dembo and Weissman prove a coding theorem for families of distortion constraints and apply it to minimax redundancy in noisy source coding \cite[Theorem~1]{dembo2003}. The simultaneous-coding step below uses this established principle; the additional structure here is the calibrated decision-regret source, the full-evidence benchmark, and the finite-information boundaries. De Heide characterizes exact preservation of all later decisions in finite experiments through sufficiency and proves an additive acquisition law for independent Gaussian streams \cite[Theorems~1 and~3]{deheide2026}. Our boundary results concern approximate preservation on standard Borel experiments and distinguish a fixed action-count bound from all finite menus jointly.

Nayebi derives predictive distinctions required by low average regret on structured decision families \cite{nayebi2026}. Here the decision-regret state is a sufficient mathematical source for the common coding problem, with physical implementation determined by the preparation interface. Furutanpey and Bunte compare finite-task coding rates under prescribed observations and permitted prediction tuples \cite[Sections~3--4]{furutanpey2026}. Those interface distinctions are retained here, while the compact-family reduction and the collective-spectrum separation address different questions.

Invariant predictive compression preserves downstream uses specified through transformations~\cite{dubois}. Hu et al. give dual-VC communication bounds, greedy selection of Boolean tasks, and algorithms for learning universal features~\cite{hu}. Our active-task certificate gives a computable bound within a declared convex design problem, with task-search costs accounted for explicitly. Gaussian multitask information-bottleneck theory gives rate-dependent allocation laws and task-priority reversals \cite{stiles}. Benedek and Itai relate fixed-distribution learnability to finite disagreement covers \cite{benedek}; the obstruction below is determined by the closure of losses relative to their full-evidence benchmarks.

These foundations support the common-representation analysis that follows. Their focus is the common representation constraint and the ordering in \cref{fig:ordering}. The contribution is a connected set of exact characterizations: a calibrated decision-state reduction, finite-information boundaries, and a separation showing that even the high-rate exponent depends on collective task structure beyond pairwise geometry. Statistical discovery and computational accessibility are then treated as additional, explicitly specified questions. The hypotheses and resource criteria are stated separately whenever the model changes. The exact parity law gives a sharp simultaneous prediction criterion for a finite-alphabet source. Its proof combines classical moment bounds with Fourier predictability and posterior entropy. The nested-resolution result in \cref{thm:nested} gives an explicit rate formula for a balanced hierarchy whose refinements and task weights are specified.

\section{Evidence, retained state, and future purpose}
\label{sec:model}
Let $\Theta$ and $Y$ be standard Borel random variables with a fixed joint law. The source $\Theta$ describes the world; $Y$ is the full preparation evidence. For sequential acquisition, let $\HH_n\uparrow\sigma(Y)$ modulo null sets. A task $t$ consists of a finite action set $\mathcal A_t$ and measurable losses
\[
0\le \ell_t(a,\theta)\le1.
\]
The task may depend on the public experiment and the coding protocol, but is fixed independently of the realized source, observations, retained message, and random seed. The decoder receives $t$ only after preparation.

For any information variable $V$, define
\begin{equation}
 b_V(t)=\E\min_{a\in\mathcal A_t}\E[\ell_t(a,\Theta)\mid V].
 \label{eq:bayes}
\end{equation}
A retained state $U$ is generated from $Y$ through a possibly randomized channel. Its regret is
\begin{equation}
 \Delta_t(U)=b_U(t)-b_Y(t),\qquad
 \Delta_{\TT}(U)=\sup_{t\in\TT}\Delta_t(U).
 \label{eq:regret}
\end{equation}
Conditional Jensen's inequality gives $\Delta_t(U)\ge0$. Independent public randomness is included in $U$ whenever the decoder can access it. In a hard-message model, $U=(A,R)$ with $R$ independent of $(\Theta,Y)$ and $A$ taking at most $2^M$ values for each realized $R$; then $I(Y;A,R)\le M$.

We use logarithms to base two for information and memory. KL divergences in Gaussian estimation formulas are explicitly stated in nats. Total variation is normalized to lie in $[0,1]$ for probability measures.

\begin{definition}[Future-task information]
The information needed to retain full-evidence decision value to accuracy $D$ is
\begin{equation}
 \RR_{\TT}(D)=
 \inf_{P_{U\mid Y}}\left\{I(Y;U):\Delta_{\TT}(U)\le D\right\}.
 \label{eq:rate}
\end{equation}
An infeasible infimum is $+\infty$. The finite-summary floor is
\begin{equation}
 \Gamma_{\TT}=\inf_{\substack{n<\infty\\A\text{ finite, }\HH_n\text{-measurable}}}
 \Delta_{\TT}(A).
 \label{eq:floor}
\end{equation}
\end{definition}

The information budget in \eqref{eq:rate} has a fixed-length operational interpretation for independent repetitions, proved in \cref{thm:operational}. One-shot message cardinality is measured separately. Finite-sample minimax learning later uses a different criterion, explicitly defined in \cref{sec:discovery}.

Let $\Pi=\mathcal L(\Theta\mid Y)$ be the full-evidence posterior. Replacing $P_{U\mid Y}$ by its conditional average given $\Pi$ preserves the joint law of $(\Theta,U)$ and cannot increase the relevant information cost. Indeed, the new channel has $I(Y;U)=I(\Pi;U)$, while data processing bounds this by the old $I(Y;U)$. Thus \eqref{eq:rate} can equivalently optimize channels from the posterior. The next section makes a further reduction.

\section{The information that decisions actually require}
\label{sec:state}
For a task with actions $[d]=\{1,\ldots,d\}$, write
\begin{align}
 L_t(Y,a)&=\E[\ell_t(a,\Theta)\mid Y],&
 B_t(Y)&=\min_b L_t(Y,b),\nonumber\\
 X(t,a)&=L_t(Y,a)-B_t(Y).\label{eq:canonical-state}
\end{align}
The random function $X$ is the joint decision-regret state. It records all posterior action comparisons simultaneously. Each coordinate lies in $[0,1]$, and $\min_aX(t,a)=0$.

\paragraph{Compact operational setting.}
In this section, the tasks are indexed by a compact metric space $T$, all use the common finite action set $[d]$, and $X$ is a Borel random element of $C(T\times[d])$ supported on a compact set $\mathcal E$ in the uniform norm. Padding with duplicate actions allows any common upper bound on action count. The assumptions hold, for example, when the posterior has compact metric support and the continuous excess-loss profiles form a compact family in the uniform topology. The compactness condition applies to the decision-loss profiles in the uniform topology.

For $x\in C(T\times[d])$, put $g_t(x)=\min_a x(t,a)$. Since evaluation is continuous and the state is bounded, its conditional expectations are well-defined as Banach-valued conditional expectations.

\begin{theorem}[Canonical calibrated-state reduction]
\label{thm:canonical}
For every retained state $U$, let $Z_0=\E[X\mid U]$. There is a channel $X\to Z$ having the same joint law of $(X,Z)$ as $(X,Z_0)$ and satisfying
\begin{equation}
 \E[X\mid Z]=Z,\qquad
 \Delta_t(Z)=\Delta_t(U)=\E g_t(Z),\qquad
 I(Y;Z)=I(X;Z)\le I(Y;U).
 \label{eq:canonical-reduction}
\end{equation}
Consequently,
\begin{equation}
 \boxed{
 \RR_{\TT}(D)=\inf\left\{I(X;Z):
 \E[X\mid Z]=Z,\quad \sup_{t\in T}\E g_t(Z)\le D\right\}.}
 \label{eq:canonical-curve}
\end{equation}
The entire curve depends only on the joint law of $X$. If $I(Y;U)<\infty$, the information discarded by this reduction is exactly
\begin{equation}
 I(Y;U)-I(X;Z)
 =I(Y;U\mid X)+I(X;U\mid Z_0).
 \label{eq:waste}
\end{equation}
\end{theorem}

\begin{proof}
The task-wise baseline cancels:
\[
\Delta_t(U)=\E\min_a\E[X(t,a)\mid U]=\E g_t(Z_0).
\]
The tower property gives $\E[X\mid Z_0]=Z_0$. Disintegrate the joint law of $(X,Z_0)$ and regenerate $Z$ through its conditional law given $X$. This preserves all displayed risks and calibration. The Markov relation $Y\to X\to Z$ gives $I(Y;Z)=I(X;Z)$, and data processing gives $I(X;Z)=I(X;Z_0)\le I(Y;U)$. Conversely, every calibrated channel in \eqref{eq:canonical-curve} is an admissible retained representation. Finally, apply the information chain rule first to $X$, a function of $Y$, and then to $Z_0$, a function of $U$. This proves \eqref{eq:waste}.
\end{proof}

The two terms in \eqref{eq:waste} describe different redundancies. The first retains evidence distinctions invisible to all declared task regrets. The second retains information about the decision state beyond its conditional mean, although that mean already determines every future Bayes risk. The theorem identifies a sufficient statistical description. Computational realization and the choice among statistically equivalent descriptions depend on the application.

\paragraph{A one-bit example.}
Suppose full evidence reveals independent fair bits $(B,C)$, but the only future task predicts $B$. Let $U=(C,B\oplus N)$ with $N\sim\Ber(q)$ independent. Both $U$ and the calibrated forecast of $B$ have error $q$. Their information costs are $2-h_2(q)$ and $1-h_2(q)$, respectively. Exactly one bit is retained without serving any possible task.

\subsection{An operational theorem for an infinite task family}
Consider $n$ independent copies of the entire experiment. Preparation encodes $Y^n$ into a fixed-length message $J_n$ before the task is known. One task $t$ is then announced and used at every coordinate. The code's distortion is its largest expected average regret over such tasks:
\begin{equation}
 D_n=\sup_t\frac1n\sum_{i=1}^n
 \left\{\E\ell_t(\widehat a_{t,i}(J_n),\Theta_i)-b_Y(t)\right\}.
 \label{eq:block-distortion}
\end{equation}
An independent shared seed may be available at both ends. The task is independent of it. Let $R_{\mathrm{op}}(D)$ be the infimum of asymptotic fixed-length rates among code sequences with $\limsup_nD_n\le D$.

\begin{theorem}[Future-task coding and least-favorable purposes]
\label{thm:operational}
Under the compact operational assumptions,
\begin{equation}
 R_{\mathrm{op}}(D)=\RR_{\TT}(D).
 \label{eq:operational-equality}
\end{equation}
Let $\nu=\mathcal L(X)$, and let $\mathcal M(\nu)$ consist of probability measures $\mu$ on $\PP(\mathcal E)$ with barycenter $\nu$. For $q\in\PP(\mathcal E)$ define
\[
 c(q)=\KL(q\Vert\nu),\qquad
 \varphi_t(q)=\min_a\int x(t,a)\,q(dx),
\]
with KL measured in bits. Then
\begin{equation}
 \RR_{\TT}(D)=
 \min_{\mu\in\mathcal M(\nu)}
 \left\{\int c(q)\,\mu(dq):
 \sup_{t\in T}\int\varphi_t(q)\,\mu(dq)\le D\right\}.
 \label{eq:splitting}
\end{equation}
If a finite-information split is uniformly strictly feasible, then
\begin{align}
 \RR_{\TT}(D)=\sup_{\beta\ge0,\,\lambda\in\PP(T)}
 \bigg\{-\beta D+\inf_{\mu\in\mathcal M(\nu)}
 \int\!\left[c(q)+\beta\int_T\varphi_t(q)\lambda(dt)\right]\mu(dq)\bigg\}.
 \label{eq:dual}
\end{align}
The multiplier measure may be taken on the compact closure of task profiles. It supports a least-favorable mixture in the Lagrangian sense.
\end{theorem}

The posterior-splitting formula follows by conditioning $X$ on the retained state. The converse to coding single-letterizes the common message; achievability uses finite source and task approximations followed by finite-alphabet multiple-distortion coding. Compactness makes the approximation uniform across the entire task family. Complete proofs, including the endpoint convention in \eqref{eq:operational-equality}, appear in \cref{app:operational}.

The operational theorem codes independent environments in blocks, with one task chosen independently of the data and applied throughout each block. Its rate measures amortized fixed-length coding; one-shot message cardinality and code-construction complexity are separate resources. Compact uniform task profiles imply $\Gamma_{\TT}=0$: quantizing $\mathcal E$ to any positive uniform accuracy yields a finite code. The next section allows broader task classes and identifies their finite-information floors. The coding theorem retains its stated compactness assumptions.

\section{The boundary of finite usefulness}
\label{sec:boundary}
Finite retention can fail even when full evidence answers every individual request. The obstruction depends on whether the number of available actions is uniformly bounded across tasks.

\subsection{A fixed finite action count}
Fix $d<\infty$. Identify a task with its loss vector $\ell\in L^\infty(P_\Theta)^d$ and its benchmark $c=b_Y(\ell)$. Let
\[
\mathcal B_{\TT}=\{(\ell_t,b_Y(t)):t\in\TT\},
\]
and take its closure in the product of the weak-star loss topology and the ordinary topology on $[0,1]$.

\begin{theorem}[Finite-information boundary]
\label{thm:closure}
For every nonempty fixed-$d$ family,
\begin{equation}
 \boxed{\Gamma_{\TT}=
 \max_{(\ell,c)\in\overline{\mathcal B}_{\TT}}
 \{b_Y(\ell)-c\}
 =\inf\{D:\RR_{\TT}(D)<\infty\}.}
 \label{eq:boundary}
\end{equation}
The first infimum is unchanged if finite summaries are replaced by any channel $\Theta\to Y\to U$ whose joint law satisfies $P_{\Theta,U}\ll P_\Theta\otimes P_U$. Finite mutual information is sufficient for this domination. The boundary $D=\Gamma_{\TT}$ is characterized as an infimum; an attaining finite summary may require additional assumptions.
\end{theorem}

A fixed finite summary has a weak-star continuous Bayes-risk functional. Full-evidence Bayes risk is upper semicontinuous. These two facts give the lower bound. For the upper bound, choose a finite Bayes-action summary near each point of the compact task--benchmark closure, take a finite subcover, and retain the tuple of its summaries. The finite observation-prefix requirement follows from martingale approximation. The complete argument is in \cref{app:closure}.

The benchmark must travel with the loss. A sequence of individually easy tasks can converge weak-star to a task whose full-evidence Bayes risk is positive, while its benchmarks converge to zero. That discrepancy is exactly what \eqref{eq:boundary} records.

\subsection{All finite action menus at once}
The next guarantee requires one retained state to serve every finite action count. The preceding result permits preparation for a fixed $d$; the present result requires one encoding to serve all finite action counts. The cost of describing and evaluating the supplied menu is not part of this information budget.

Write
\begin{equation}
 P_{\Theta,Y}=P_{\mathrm{ac}}+P_{\mathrm{s}},\qquad
 P_{\mathrm{ac}}\ll Q,\quad P_{\mathrm{s}}\perp Q,\quad
 Q=P_\Theta\otimes P_Y,
 \label{eq:lebesgue}
\end{equation}
and let $s=P_{\mathrm{s}}(\Theta\text{-space}\times Y\text{-space})$. Let $\TT_{\mathrm{all}}$ contain every measurable $[0,1]$-valued loss vector with any finite action count.

\begin{theorem}[The singular-information obstruction]
\label{thm:singular}
For standard Borel source and evidence,
\begin{equation}
 \boxed{
 \inf_{A\text{ finite prefix summary}}\Delta_{\TT_{\mathrm{all}}}(A)
 =s
 =\inf\{D:\RR_{\TT_{\mathrm{all}}}(D)<\infty\}.}
 \label{eq:singular-law}
\end{equation}
Every retained $U$ with $P_{\Theta,U}\ll P_\Theta\otimes P_U$ satisfies $\Delta_{\TT_{\mathrm{all}}}(U)\ge s$; the infimum over this class equals $s$. The same lower bound and infimum hold when the class is restricted to finite $I(\Theta;U)$. One countable sequence of finite-action tasks, depending only on the experiment, witnesses the lower bound for all such retained states.
\end{theorem}

The proof has a concrete interpretation. On the singular part of the experiment, full evidence can select a source set of arbitrarily small prior probability and nevertheless contain the true source with probability approaching $s$. A dominated retained state cannot do so: its probability of selecting any such small-prior set tends to zero. Each selection problem has finitely many actions, although their number grows as the sets shrink. Conversely, the absolutely continuous part can be approximated in $L^1$ by a finite evidence partition, preserving every bounded decision at once. \Cref{app:singular} gives both constructions.

Exact revelation of a nonatomic source has $s=1$; exact revelation of a countably atomic source has $s=0$. The floor is determined by singular mass, and sources with infinite mutual information can have different floors. The theorem concerns decision value relative to full evidence, and allows unlimited but finite action menus. The value of the floor for a restricted task class depends on that class and its losses.

\subsection{Why the action count changes the answer}
Let $Tf=\E[f(\Theta)\mid Y]$ and regard $T$ as an operator from $L^\infty(P_\Theta)$ to $L^2(P_Y)$. For a finite statistic $A=A(Y)$, set
\[
\varepsilon(A)=\sup_{\norm f_\infty\le1}\norm{Tf-\E[Tf\mid A]}_2.
\]
Let $\Delta_d(A)$ be worst regret over all bounded tasks with at most $d$ actions.

\begin{proposition}[Fixed menus and operator compactness]
\label{prop:compact}
For every finite deterministic statistic,
\begin{equation}
 \frac{\varepsilon(A)^2}{8}\le\Delta_2(A)\le\frac{\varepsilon(A)}2,
 \qquad \Delta_d(A)\le\frac{\sqrt d}{2}\varepsilon(A).
 \label{eq:compact-bounds}
\end{equation}
Zero finite-summary floor for binary decisions, zero floor for every fixed finite $d$, and compactness of $T:L^\infty(P_\Theta)\to L^2(P_Y)$ are equivalent.
\end{proposition}

These conditions can hold even when $s=1$, as the following example shows. Let the source consist of independent fair bits $B_j$, and independently reveal bit $j$ with probability $p_j=1/j$. Full evidence reveals infinitely many coordinates almost surely, so its joint law is entirely singular. Yet the conditional-expectation operator is compact: after retaining the first $m$ observations, its unit-ball residual norm is at most $1/\sqrt{m+1}$. Thus
\begin{equation}
 \sup_{d<\infty}\inf_A\Delta_d(A)=0,
 \qquad
 \inf_A\sup_{d<\infty}\Delta_d(A)=1.
 \label{eq:quantifier-separation}
\end{equation}
The preparation may depend on a declared action-count bound in the first expression; it must serve every finite bound in the second. The quantifier order determines whether preparation may depend on the action-count bound.

\section{The collective structure of purposes}
\label{sec:collective}
The canonical state records the joint arrangement of tasks. We now show why a pointwise pairwise metric cannot replace it.

\subsection{An elementary separation}
Let $d\ge3$ and let $\Theta=(B,J)$ be uniform on $\{0,1\}\times[2^{d-1}]$. The bit $B$, together with the $(d-1)$-bit binary encoding of $J-1$, gives $d$ independent fair bits. Compare two future-task families under zero--one loss. The coordinate family requests any one of those bits. The shared family requests
\[
f_j(B,J)=B\oplus\ind\{J=j\},\qquad j\in[2^{d-1}].
\]
Every individual answer is fair. Both families separate every pair of distinct source states, so their pointwise supremum disagreement metric is the same discrete metric.

Nevertheless, for $0<D<1/2$,
\begin{equation}
 \RR_{\mathrm{coord}}(D)=d[1-h_2(D)],\qquad
 \RR_{\mathrm{shared}}(D)\longrightarrow1-h_2(D)
 \quad(d\to\infty).
 \label{eq:finite-separation}
\end{equation}
For the upper bound on the shared family, transmit a noisy version of $B$ and omit $J$. Each individual exception has probability $2^{1-d}$. A lower bound follows by applying binary Fano to any one fair task. At zero error both families require $d$ bits, since they separate all states. Positive tolerance makes the common structure visible. The exact bounds are proved in \cref{app:metric}.

\begin{figure}[t]
\centering\includegraphics[width=.75\linewidth]{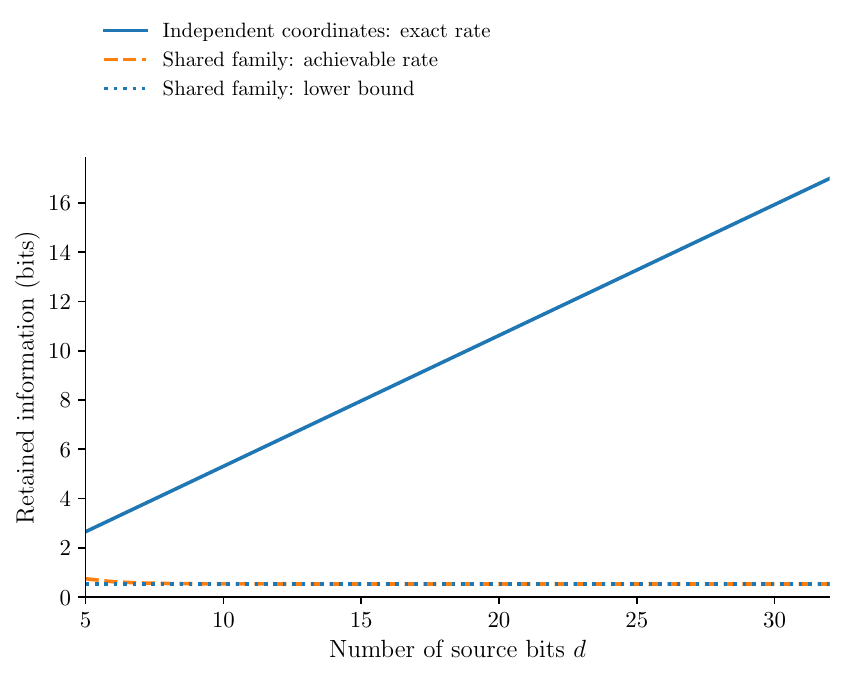}
\caption{The same source and pairwise task metric can have different common-retention costs. At $D=0.1$, the coordinate rate is exact. The shared-family curve is an explicit achievable information rate; its lower bound is $1-h_2(D)$. All individual tasks have the same fair binary answer distribution. All curves are evaluated from analytic bounds or constructions.}
\label{fig:metric}
\end{figure}

\subsection{The exponent can change while the metric stays fixed}
The finite example separates positive-distortion rates, but its zero-error endpoint is still the same for both families. To test whether the collective arrangement can change the leading rate as accuracy increases, we need distinctions at successively finer scales. The next construction supplies them while preserving the source, the pairwise metric, and every individual-task curve at its scale.

Fix $0<a<1$, and let $\Theta=(B,X_1,X_2,\ldots)$ consist of independent fair bits. If two sequences $x,x'$ have $L$ initial bits in common, set $d_X(x,x')=a^L$ and put
\[
Q=\frac{\log2}{\log(1/a)}.
\]
The metric space of sequences is compact and has covering dimension $Q$ in the metric-entropy sense. For depth $j$, let $J_j$ be the length-$j$ prefix, with $N_j=2^j$ possible values. For $0\le\beta\le1$, set
\[
\alpha_j=a^{j-1},\qquad s_j=\max\{1,\lfloor N_j^\beta/2\rfloor\}.
\]
A task chooses $S\subseteq[N_j]$ with $1\le|S|\le s_j$, asks for $B\oplus\ind\{J_j\in S\}$, and charges weighted binary error $\alpha_j^2$. Include the task requesting $B$ with unit weight and a null task. Denote the resulting family by $\TT_\beta$.

Every nonnull task has a fair binary answer. For a task of weight $w$, its known-task information curve is $1-h_2(D/w)$ on $0\le D\le w/2$, and zero beyond that range. The response supremum metric, with response amplitude $\alpha_j$, is independent of $\beta$:
\begin{equation}
 \widetilde d((b,x),(b',x'))=
 \begin{cases}1,&b\ne b',\\d_X(x,x'),&b=b'.\end{cases}
 \label{eq:common-metric}
\end{equation}
The corresponding supremum metric on loss vectors is $\widetilde d^{\,2}$.

\begin{theorem}[Collective composition spectrum]
\label{thm:spectrum}
Let $V_\beta(K)$ be the least worst-task expected loss using at most $K$ message values, with arbitrary stochastic encoding and independent shared randomness. Define
\[
\rho_\beta=1-\beta+\frac2Q.
\]
There are positive constants depending only on $a,\beta$ such that
\begin{equation}
 cK^{-\rho_\beta}\le V_\beta(K)\le CK^{-\rho_\beta}.
 \label{eq:spectrum-card}
\end{equation}
Moreover,
\begin{equation}
 \boxed{\RR_{\TT_\beta}(D)
 =\frac{Q}{2+(1-\beta)Q}\log_2\frac1D+O(1),\qquad D\downarrow0.}
 \label{eq:spectrum-rate}
\end{equation}
These families have the same source law, the same pairwise response metric, and the same per-scale individual-task rate curves. No task identifies the whole source.
\end{theorem}

A question may test one small region or combine many regions before it is asked. The exponent records this collective reach. The parameter $\beta$ changes how many regions a single future question can combine while preserving the pairwise distinctions available across the family.

The proof uses a source-weighted task covariance. At one scale, for a posterior vector $p$ on $[N]$ and a uniform $s$-element subset $S$,
\begin{equation}
 \E_S[p_S(1-p_S)]
 =\frac{s(N-s)}{N(N-1)}(1-\norm p_2^2).
 \label{eq:subset-identity}
\end{equation}
A $K$-valued message leaves average collision probability at most $K/N$. This gives the cardinality converse. To obtain the exact logarithmic coefficient for mutual information, one must use all nested scales: the task guarantees force successively smaller errors on coarser prefix bits, and their conditional entropies have a summable tail. \Cref{app:spectrum} provides the complete argument.

\begin{figure}[t]
\centering\includegraphics[width=.75\linewidth]{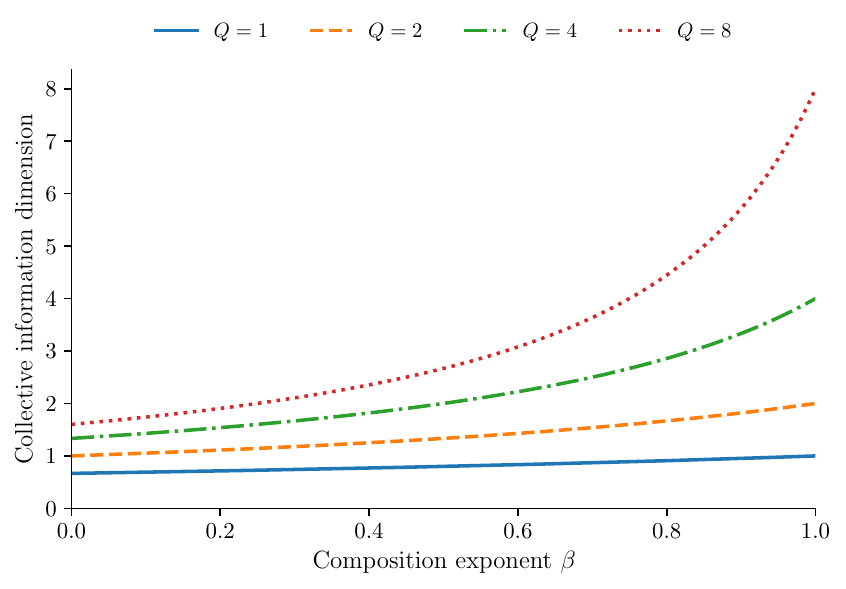}
\caption{Collective information dimension at fixed pairwise geometry. The plotted value is $2Q/[2+(1-\beta)Q]$, twice the coefficient of $\log_2(1/D)$ in \cref{thm:spectrum}. Each curve keeps the source and its pairwise metric fixed. The changing exponent is caused only by the number of regions that one task can combine.}
\label{fig:composition}
\end{figure}

\paragraph{Covariance and the shape of the information curve.}
Second-order task statistics can miss both the amount of common information and its dependence on tolerated error. Independent coordinate questions and all binary parity questions have pairwise uncorrelated answers, but their joint dependence produces different optimal codes.

\begin{proposition}[Exact coordinate and parity retention]
\label{prop:coordinate-parity}
Let $k\ge1$, $N=2^k-1$, and let full evidence reveal $N$ independent fair bits. The coordinate family requests each source bit; the parity family requests every nonzero binary parity of the first $k$ bits. Both signed answer vectors have mean zero and covariance $I_N$. For every $0\le D\le1/2$, under zero--one loss with the task fixed independently of the retained state, their information rates are
\begin{align}
 R_{\rm coord}(D)&=N[1-h_2(D)],\label{eq:coordinate-parity-coord}\\
 R_{\rm par}(D)&=k-h_2(\varepsilon_k)-\varepsilon_k\log_2N,
 \qquad \varepsilon_k=2D(1-2^{-k}),\label{eq:coordinate-parity-bounds}
\end{align}
\end{proposition}

\begin{corollary*}[Uniform-average and worst-task parity errors]
For the uniform binary source and all nonzero parity tasks, replacing the worst-task error constraint by the uniformly averaged task-error constraint leaves the rate unchanged:
\[
 R_k^{\rm avg}(D)=R_k^{\rm par}(D)
 =k-h_2(\varepsilon_k)-\varepsilon_k\log_2(2^k-1),
 \qquad 0\le D\le\tfrac12,
\]
where $\varepsilon_k=2D(1-2^{-k})$. The average is over the $2^k-1$ nonzero tasks, independently of the source, message, and shared seed.
\end{corollary*}
The converse in Appendix~\ref{app:parity-retention} uses only this average error, while the attaining channel gives every task error exactly $D$. Thus average and uniform guarantees have the same optimal information cost for this family.

The coordinate equality follows from binary Fano and independent binary test channels. For parities, let $W$ contain the first $k$ source bits and retain $U=W\oplus TV$, with independent $T\sim\Ber(2D)$ and $V\sim\Unif(\mathbb F_2^k)$. Neither $T$ nor $V$ is separately retained. The channel leaves the source unchanged or replaces it by a uniform value, without recording which occurred. Every later parity then has error $D$.

The converse is an entropy extremal statement. For a posterior $p$ on $\mathbb F_2^k$, let $\beta(p)$ be the average magnitude of its nontrivial Walsh coefficients and put $q_\beta=[1+(2^k-1)\beta]/2^k$. Then
\begin{equation}
 H(p)\le h_2(q_{\beta(p)})+(1-q_{\beta(p)})\log_2(2^k-1).
 \label{eq:parity-envelope-main}
\end{equation}
Equality is attained by a point mass mixed with the uniform distribution. Predicting every parity to error $D$ forces the average posterior value of $\beta$ to be at least $1-2D$. Concavity of the envelope then gives the exact information lower bound. Appendices~\ref{app:covariance-limits}--\ref{app:parity-refinement} prove the inequality and the coding result, using Fourier orthogonality, a classical bounded-variance exponential-moment bound, and entropy duality \cite{odonnell,bennett1962,seneta1982}.

Per bit of the exact common description, the coordinate curve is $1-h_2(D)$ and the parity curve approaches $1-2D$. Their difference is qualitative: at fixed $0<D<1/2$, the gain over explicit erasure is $N[h_2(D)-2D]$ for coordinates, but only $h_2(2D)+o(1)$ bits for parities. In fact,
\begin{equation}
 R_{\rm par}(D)=(1-2D)k-h_2(2D)+O_D(k2^{-k}).
 \label{eq:parity-symmetric-asymptotic}
\end{equation}
The exact rate applies to all nonzero binary parities of a uniform source under zero--one loss. The covariance comparison preserves unconditional task covariance; \cref{thm:spectrum} additionally preserves the pointwise source metric.

The test channel outputs a $k$-bit vector on each instance. Its mutual-information cost is attained asymptotically as a fixed-length rate through iid block coding; the single-letter output has $k$ bits. The converse covers arbitrary randomized representations and task-dependent decoders. Finite-block message size and codebook-search complexity are separate questions.

\paragraph{What unweighted dual VC dimension sees.}
Discard the weights of the binary task labels in $\TT_\beta$. For every $\beta>0$, the resulting class has infinite dual VC dimension. Indeed, choose a depth with $2^d\le N_j$ and $2^{d-1}\le s_j$. Label $2^d$ prefixes by the patterns in $\{0,1\}^d$, and let the $i$th of $d$ tasks select the prefixes whose $i$th pattern bit is one. With $B=0$, every label pattern occurs. Since $s_j\to\infty$, $d$ is arbitrary. Thus the global $O(d^*\log(1/\varepsilon))$ bound of Hu et al.\ \cite{hu} is vacuous for this unweighted class, whereas \eqref{eq:spectrum-rate} is finite. The distinction is loss weighting across resolutions: fine questions permit larger unweighted error at a fixed weighted tolerance. The comparison concerns the global unweighted dual-VC bound. Refined dual-growth and distribution-specific constructions may yield useful bounds after truncation.

\subsection{A general law for nested resolutions}
\label{sec:nested}
The power law in \cref{thm:spectrum} is one instance of a more general retention rule. At each resolution, three quantities matter: how many regions exist, how many a single question can combine, and the loss attached to getting that question wrong.

Let $W_j$ be independent and uniform on $[b_j]$, where $2\le b_j\le b_*<\infty$, and let $B$ be an independent fair bit. Put $J_j=(W_1,\ldots,W_j)$ and $N_j=\prod_{i=1}^j b_i$, with $N_0=1$. Preparation observes $(B,W_1,W_2,\ldots)$. At level $j$, the task family contains every question $B\oplus\ind\{J_j\in S\}$ with $1\le|S|\le s_j\le N_j/2$, charged at binary-error weight $w_j\in(0,1]$. Include the unit-weight mask task and the null task, and assume $w_j\to0$.

The level cost is $w_js_j/N_j$: after storing $B$ but omitting the region, answering $B$ incurs at most this loss. To allow bounded oscillations caused by integer reaches, let $c_j$ be a positive nonincreasing profile with
\begin{equation}
 a_-c_j\le \frac{w_js_j}{N_j}\le a_+c_j,
 \qquad c_j\downarrow0,
 \qquad c_{j+1}\ge\eta c_j,
 \label{eq:nested-profile}
\end{equation}
for fixed $a_-,a_+>0$ and $0<\eta\le1$. One may take $c_j=w_js_j/N_j$ whenever that sequence is regular. Write $h_2$ for binary entropy and assume
\begin{equation}
 \mathcal E_c:=\sup_{n\ge1}\sum_{j=1}^n
 h_2\!\left(\frac{c_n}{2c_j}\right)<\infty.
 \label{eq:nested-entropy}
\end{equation}
This condition has an equivalent resolution-scale interpretation:
\begin{equation}
 \mathcal E_c<\infty
 \quad\Longleftrightarrow\quad
 \exists L\ge1\ \ \forall j\ge1:\quad c_{j+L}\le\tfrac12c_j.
 \label{eq:block-equivalence}
\end{equation}
Thus every fixed number of refinement steps must reduce unresolved cost by a fixed factor. The sequence need not be geometric at each step, and branching may vary. Appendix~\ref{app:decay-regimes} proves the equivalence using only monotonicity of $c_j$; the adjacent-ratio condition serves a separate role in the coding bounds.

\begin{theorem}[Nested-resolution retention law]
\label{thm:nested}
For the preceding task family $\TT_{\rm nest}$, put
\[
 j(D)=\min\{j\ge1:c_j\le D\},\qquad
 n(K)=\max\{n\ge0:2N_n\le K\}.
\]
For every integer $K\ge2$, the optimal hard-message loss satisfies
\begin{equation}
 \frac{a_-\eta^2}{4}\,c_{n(K)+1}
 \le V_{\rm nest}(K)\le a_+c_{n(K)+1}.
 \label{eq:nested-cardinality}
\end{equation}
For mutual information, and for the iid operational criterion of \cref{thm:operational},
\begin{equation}
 \boxed{\RR_{\TT_{\rm nest}}(D)=\log_2N_{j(D)}+O(1),\qquad D\downarrow0.}
 \label{eq:nested-rate}
\end{equation}
The constants depend only on the displayed regularity parameters. The converses allow arbitrary stochastic encoders and independent shared randomness. Retaining the mask bit and a sufficiently long prefix attains the rate within a constant number of bits.
\end{theorem}

The upper construction resolves coarse questions exactly and accepts the bounded cost of finer questions. The converse uses all resolutions: small task errors force recovery of successive refinement digits, whose remaining conditional entropies have a uniformly bounded sum. The additive $O(1)$ is stronger than an exponent match: the explicit mask--prefix description costs at most a constant number of bits more than any retained channel. \Cref{app:nested} gives both proofs, including the bit for $B$ and the constant-error changes in the crossing index $j(D)$.

For \cref{thm:spectrum}, $b_j=2$ and $w_js_j/N_j\asymp2^{-\rho_\beta j}$, so the new law recovers its coefficient. It also resolves corrections beyond a power law. For $b_j=2$, $s_j=N_j/2$, and $w_j=2^{-\rho j}(j+1)^{-\zeta}$ with $\rho>0$ and $\zeta\ge0$,
\begin{equation}
 \RR(D)=\frac1\rho\log_2\frac1D
 -\frac\zeta\rho\log_2\log_2\frac1D+O(1).
 \label{eq:nested-loglog}
\end{equation}
The log-log term separates hierarchies with the same leading information dimension. The inverse cost profile contains finer information than that dimension alone.

\paragraph{Outside uniform block contraction.}
A different rate can arise when fine questions lose importance more slowly. With binary refinements, $s_j=N_j/2$, and $w_j=j^{-p}$ for fixed $p>0$, the level cost is $c_j=\tfrac12j^{-p}$ and
\begin{equation}
 \RR(D)=\Theta(D^{-1/p}),\qquad D\downarrow0.
 \label{eq:polynomial-rate}
\end{equation}
Its rate divided by $\log_2(1/D)$ diverges. A profile with increasingly long flat intervals has $\mathcal E_c=\infty$ while still obeying $\RR(D)=\Theta(\log_2(1/D))$. Appendix~\ref{app:decay-regimes} supplies both constructions and converses. Uniform block contraction characterizes the entropy-tail hypothesis; the polynomial and plateau examples exhibit distinct rates beyond that hypothesis.

The hierarchy, its probability balance, and the task weights are declared inputs. The result counts source-dependent retained information; task descriptions and membership evaluation are separate costs. The coding law applies to this declared hierarchical class; learning a hierarchy from raw observations is treated as a separate statistical problem.

\subsection{When geometric dimension is sufficient}
Classical rate--distortion dimension relates logarithmic rate growth to geometric dimension \cite{kawabata}. In the present setting, a regular $k$-dimensional metric gives coefficient $k/2$ when tasks collectively control that geometry in a distributionally appropriate way. A sufficient local condition is a finite diagnostic collection $t_1,\ldots,t_L$ with
\[
 c\norm{z-z'}^2\le\sum_{\ell=1}^L
 \norm{q_{z,t_\ell}-q_{z',t_\ell}}_2^2
 \le C\norm{z-z'}^2,
\]
together with a global task-Lipschitz bound, compact regular parameter support, and a nondegenerate source density. Covering gives an upper rate of $(k/2)\log_2(1/D)+O(1)$; reconstruction from the diagnostic tuple and an entropy bound give the reverse inequality. The constants are fixed with the geometry. Dimension-uniform remainders require corresponding uniform assumptions. \Cref{thm:geometric} states this precisely. \Cref{thm:spectrum} exhibits different rate exponents when only pairwise geometry and individual-task rates are fixed.

\section{Forgetting, independent purposes, and reuse}
\label{sec:reuse}
The same decision-state formulation describes how useful information is lost in stages and shared across purposes.

If $Y\to U\to V$, then for every task
\begin{equation}
 b_V(t)-b_Y(t)
 =[b_U(t)-b_Y(t)]+[b_V(t)-b_U(t)].
 \label{eq:forgetting}
\end{equation}
Every term is nonnegative. Thus successive forgetting telescopes task by task. The supremum over tasks need not telescope with equality: different stages can damage different purposes. The joint decision state retains that distinction.

Let $X_1,\ldots,X_m$ be decision states for component task families satisfying the compact operational assumptions, with curves $\RR_i$. A common representation must meet regret targets $D_i$ for every component. For additive product tasks, use weights $w_i>0$ with $\sum_iw_i=1$.

\begin{theorem}[Independent purposes and the value of reuse]
\label{thm:reuse}
If the component states are independent, then
\begin{align}
 \RR_{\parallel}(D_1,\ldots,D_m)&=\sum_i\RR_i(D_i),\label{eq:direct-sum}\\
 \RR_{\mathrm{prod}}(D)&=
 \inf_{\sum_iw_iD_i\le D}\sum_i\RR_i(D_i).\label{eq:infconv}
\end{align}
For dependent states with finite total correlation
\[
\TC(X)=\KL\!\left(P_X\middle\Vert\bigotimes_iP_{X_i}\right),
\]
let $S(D)$ denote the separate-coding expression on the right of \eqref{eq:infconv}. Then
\begin{equation}
 \max\{0,S(D)-\TC(X)\}\le\RR_{\mathrm{joint}}(D)\le S(D).
 \label{eq:reuse-bound}
\end{equation}
Whenever the quantities are finite, the exact information identity is
\begin{equation}
 \sum_i I(X_i;U)-I(X;U)=\TC(X)-\TC(X\mid U).
 \label{eq:tc-identity}
\end{equation}
\end{theorem}

For a fixed finite collection of purposes, independent requirements cannot share an information bill merely because they share a decoder. Dependent requirements can; the two-sided saving bound assumes finite total correlation. Equation \eqref{eq:tc-identity} quantifies the saving: a common representation reuses information when it explains dependence among the component decision states. Total correlation is the classical multivariate quantity introduced by Watanabe \cite{watanabe}; the theorem applies it to the common future-task criterion.

If each marginal curve has coefficient $k_i/2$ at high resolution and total correlation is finite, the joint coefficient remains $\sum_i k_i/2$. Finite dependence can change an additive information constant but cannot change that logarithmic slope. Exact copies of a continuous state have infinite total correlation and can reduce the slope. The two regimes reflect finite versus infinite total correlation and give different reuse behavior.

\section{Acquisition and retention are different resources}
\label{sec:resources}
The information curve starts from available full evidence. A preparation system also has to obtain that evidence. We give two exact models where the acquisition and retention stages can be solved together, and then a nonlinear model in which the reusable representation itself is unknown.

\subsection{An exact transition for uncertain binary evidence}
Draw $P\sim\Unif[0,1]$. Given $P$, draw a binary target $\Theta\sim\Ber(P)$ and $n$ independent preparation observations $Y_i\sim\Ber(P)$, independently of the target. The benchmark observes $P$. The encoder sees the $n$ observations and retains one of $K$ states. A later task may be any finite decision on $\Theta$ with losses in $[0,1]$.

Let $V^{\mathrm{priv}}_{n,K}$ denote optimal worst-task regret when only the message is retained, and $V^{\mathrm{pub}}_{n,K}$ the value with an independent shared seed. The task is fixed independently of every realized seed.

\begin{theorem}[Exact binary acquisition--retention transition]
\label{thm:fr-joint}
If $K_n/\sqrt n\to\kappa\in(0,\infty)$, then
\begin{equation}
 \boxed{\lim_{n\to\infty}nV^{\mathrm{priv}}_{n,K_n}
 =\frac14\coth^2\kappa,\qquad
 \lim_{n\to\infty}nV^{\mathrm{pub}}_{n,K_n}
 =\frac14\coth^2(\sqrt3\kappa).}
 \label{eq:fr-joint}
\end{equation}
The converse covers arbitrary stochastic encoders of the full acquisition history. Explicit interval codes of a randomized sufficient statistic attain the limits.
\end{theorem}

The count $S=\sum_iY_i$ is uniform on $\{0,\ldots,n\}$ and has posterior mean $(S+1)/(n+2)$. Randomize privately within its probability interval to obtain $W\sim\Unif[0,1]$ with target posterior
\begin{equation}
 v_n(w)=\frac{\lceil(n+1)w\rceil}{n+2}\quad\text{almost everywhere}.
 \label{eq:fr-step}
\end{equation}
For private coding, partition $W$ at $T_a(j/K)$, where
\begin{equation}
 T_a(u)=\frac12+\frac{\sinh(a(2u-1))}{2\sinh a},\qquad a=K/\sqrt n.
 \label{eq:fr-compander}
\end{equation}
For shared coding, use $a=\sqrt3K/\sqrt n$ and shift a uniform lattice in the cumulative coordinate before adding the endpoints. The decoder uses the actual conditional mean of the step posterior in its retained interval.

The limiting curve couples acquisition uncertainty with the allocation of retained memory states. Uncertainty left by sampling changes where the memory states should be allocated. The balance is $K\asymp\sqrt n$, or about $\tfrac12\log_2 n$ retained bits. Shared randomness changes the memory-limited constant, but both curves approach the acquisition limit $1/4$. \Cref{app:binary} proves the result through the curvature representation of binary Bayes risk, the ordered-interval reduction, and a uniform discrete-to-continuous comparison. Integral representations of binary decision risk and functional quantization are established antecedents \cite{reid,misra,varshney}.

\Needspace{3\baselineskip}
\paragraph{Why memory allocation changes with evidence.}
The binary transition illustrates a broader allocation principle. Let $P$ be a full-evidence binary posterior and $Q_n=\E[P\mid\HH_n]$. Its acquisition profile is
\[
 e_n^{\mathrm{acq}}(t)=\E(P-t)_+-\E(Q_n-t)_+.
\]
For regular continuous posterior experiments, suppose $Q_n$ has a positive $C^1$ density converging in $C^1$ to $f$, and $n e_n^{\mathrm{acq}}\to a$ uniformly. With $K/\sqrt n\to\kappa$, a positive interval point density $\lambda$ gives local compression peaks $f(t)/(8\kappa^2\lambda(t)^2)$ at the scaled-regret level. The affine decision envelope in \eqref{eq:fr-envelope} therefore allocates the remaining allowance $C h_\rho(t)-a(t)$ according to
\[
 \lambda(t)\propto\left[\frac{f(t)}{C h_\rho(t)-a(t)}\right]^{1/2},
 \qquad h_\rho(t)=\rho t+(1-\rho)(1-t),
\]\nopagebreak[4]
provided that allowance is uniformly positive; $C$ and $\rho\in[0,1]$ balance the regret target and the endpoint losses. When the posterior is supplied exactly, $a=0$ and the rule reduces to $\lambda(t)\propto[f(t)/h_\rho(t)]^{1/2}$. Memory is thus assigned according to both the frequency of a belief and the decision uncertainty left by acquisition. \Cref{app:allocation-context} derives the local rule. The finite-$n$ Bernoulli posterior above is discrete; its hyperbolic limit follows from the separate discrete-to-continuous comparison.

\begin{figure}[t]
\centering\includegraphics[width=.75\linewidth]{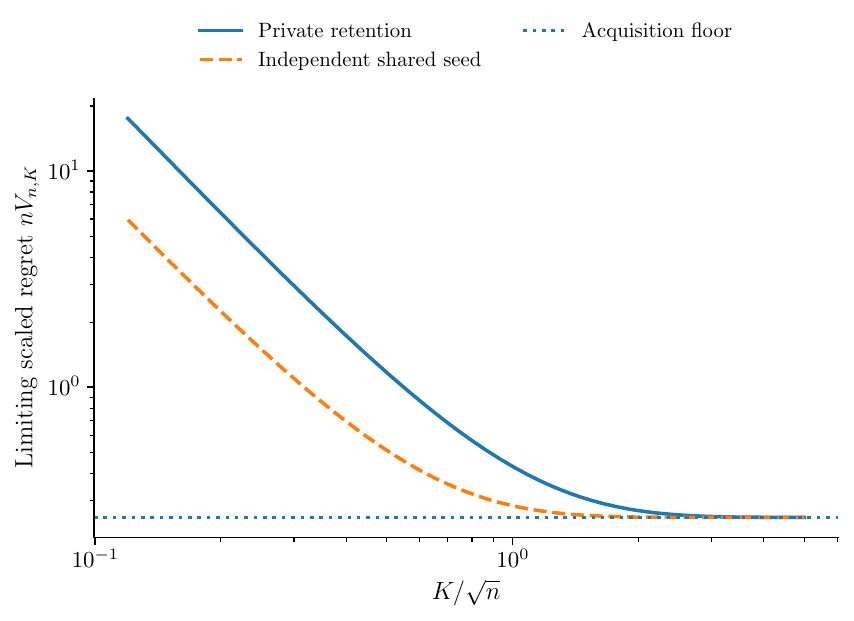}
\caption{Exact limiting binary regret versus $K/\sqrt n$. The two curves are the constants in \cref{thm:fr-joint}. Their common acquisition floor is $1/4$. Shared randomness improves the retention-limited constant under a task independent of the realized seed. The curves evaluate the proved asymptotic formulas as $K/\sqrt n$ varies.}
\label{fig:binary}
\end{figure}

\subsection{An exact Gaussian law for competing tasks}
Let a Gaussian hidden state admit the decomposition
\[
\Theta=Z+E,\qquad Z=\E[\Theta\mid Y],\qquad
 Z\sim N(0,A),\quad E\sim N(0,S),\quad E\perp Y,
\]
where $Y$ is jointly Gaussian evidence and $A\succ0$. A task asks for $B\Theta$ under squared error, with $G=B^\top B$ belonging to a compact family $\mathcal G\subset\mathbb S_+^k$. This subsection uses squared losses and a mutual-information budget; its distortion includes acquisition error relative to the hidden state.

\begin{proposition}[Collective Gaussian benchmark]
\label{prop:gaussian}
Under $I(Y;U)\le M$, the exact worst-task distortion is
\begin{equation}
 D_{\mathcal G}(M)=
 \min_{\substack{0\prec K\preceq A\\\frac12\log_2\det(AK^{-1})\le M}}
 \sup_{G\in\mathcal G}\tr\bigl(G(S+K)\bigr).
 \label{eq:gaussian}
\end{equation}
Every feasible $K$ is attained by a Gaussian channel from $Z$. A least-favorable task mixture exists and can be supported on at most $k(k+1)/2+1$ active tasks.
\end{proposition}

For arbitrary encoding, $K=\E\operatorname{Cov}(Z\mid U)$ is positive definite at finite rate, and Gaussian maximum entropy gives the determinant constraint. Gaussian test channels attain equality. Minimax duality and Carath\'eodory give the finite certificate. These are classical Gaussian coding ingredients \cite{chechik}; \cref{app:gaussian} gives their complete application to the present worst-task criterion.

For $m$ noisy measurements $Y_i=\Theta+\xi_i$ of $\Theta\sim N(0,I_k)$, with $\xi_i\sim N(0,\sigma^2I_k)$, and later scalar projection tasks $t^\top\Theta$, one obtains
\begin{equation}
 D^*(m,M)=\frac{\sigma^2}{m+\sigma^2}
 +\frac{m}{m+\sigma^2}2^{-2M/k}.
 \label{eq:gaussian-isotropic}
\end{equation}
Each task sees one direction; the collection sees all directions. This is an exact mutual-information formula with the usual iid amortized coding interpretation; the one-shot $M$-bit problem uses a distinct resource constraint. Its hidden state is random under a fixed prior. The next section instead fixes an unknown parameter and takes worst-parameter risk after finitely many raw samples.

\section{Discovering an unknown nonlinear representation}
\label{sec:discovery}
A coding theorem can describe a useful state without explaining how to learn it. The following model supplies an explicit statistical construction without a dominant future task.

Let
\[
 \gamma^2=\E\sin^2G=e^{-1}\sinh1,\qquad
 g(z)=\sin z/\gamma,\qquad G\sim N(0,1).
\]
For even $r\ge4$, $d\ge2$, and fixed $0<\alpha<A$ and $\sigma>0$, the unknown representation is
\begin{equation}
 F_{a,u}(x)=a\,g(u^\top x),\qquad
 \alpha\le\norm a\le A,\quad u\in S^{d-1}.
 \label{eq:ridge}
\end{equation}
The parameters $(a,u)$ and $(-a,-u)$ represent the same function. Raw preparation data are
\begin{equation}
 X_i\sim N(0,I_d),\qquad
 Y_i=F_{a,u}(X_i)+\xi_i,\qquad
 \xi_i\sim N(0,\sigma^2I_r),
 \label{eq:ridge-data}
\end{equation}
independently across $i$. After a hard $M$-bit message is fixed, a task supplies an orthogonal rank-$r/2$ projector $P$ and asks for $PF_{a,u}(X)$ on a fresh Gaussian input. Define
\begin{equation}
 \mathcal R_{m,M}=\inf_{\text{protocols}}\sup_{a,u}\sup_P
 \E_{\mathrm{train},X}\norm{D_P(U,X)-PF_{a,u}(X)}^2.
 \label{eq:minimax-ridge}
\end{equation}
The task is fixed independently of training data and seed, and the decoder may otherwise be arbitrary.

\begin{theorem}[Raw nonlinear discovery with collective future use]
\label{thm:nonlinear}
Put $k=r+d-1$. Constants depending only on $\alpha,A,\sigma$ satisfy
\begin{equation}
 \boxed{c\left[\left(1\wedge\frac{k}{m}\right)+2^{-2M/k}\right]
 \le\mathcal R_{m,M}\le
 C\left[\left(1\wedge\frac{k}{m}\right)+2^{-2M/k}\right].}
 \label{eq:ridge-law}
\end{equation}
No fixed task identifies the representation. A globally defined estimator uses the leading singular triplet of $m^{-1}\sum_iY_iX_i^\top$ and attains the acquisition term. An existential covering code attains the dimension-uniform retention term. A direct polynomial-time chart code attains the same bit exponent with a dimension-dependent prefactor.
\end{theorem}

Three identities supply the proof. With $\rho=u^\top v$,
\begin{align}
 \E[g(u^\top X)g(v^\top X)]&=\frac{\sinh\rho}{\sinh1}=:\kappa(\rho),\label{eq:sine-kernel}\\
 \norm{F_{a,u}-F_{b,v}}_{L^2}^2
 &=\norm a^2+\norm b^2-2a^\top b\,\kappa(u^\top v),\label{eq:sine-metric}\\
 \E[YX^\top]&=\frac{a u^\top}{\sqrt{\sinh1}}.\label{eq:sine-moment}
\end{align}
The Gaussian regression KL divergence in nats is the squared distance in \eqref{eq:sine-metric} divided by $2\sigma^2$. Locally, this distance is uniformly equivalent to the Euclidean distance in a $k$-dimensional parameter chart. Fano and an entropy argument give the sample and memory converses.

\begin{figure}[t]
\centering\includegraphics[width=.75\linewidth]{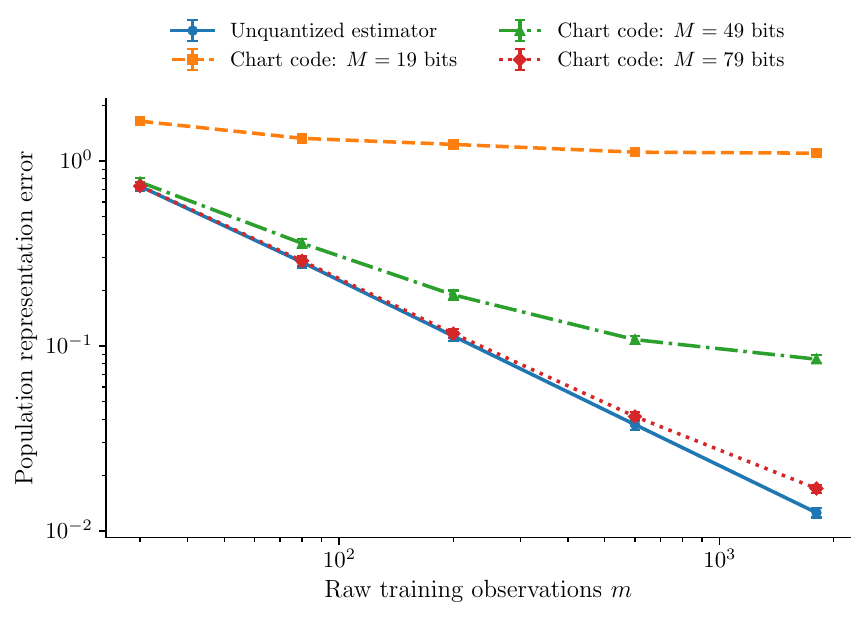}
\caption{Raw nonlinear discovery followed by finite retention. The experiment uses \eqref{eq:ridge-data}; each curve reports exact full-population representation error computed from \eqref{eq:sine-metric}, averaged over independent training experiments. This error bounds the operational worst-projector loss from above and differs from its coherent-decoder value by at most a factor two. The curves use the executable chart code with its explicit dimension-dependent overhead. Error bars are marginal standard-error intervals.}
\label{fig:nonlinear}
\end{figure}

For any fixed pair of representations, their output-difference second moment has rank at most two. A rank-$r/2$ task can contain its range and therefore recover the full pairwise $L^2$ difference. No fixed projector can do this for every pair: its kernel permits an invisible perturbation of $a$. Two fixed complementary tasks suffice for the converse analysis, where both decoders are evaluated on the same message. The deployment experiment supplies whichever single task is announced after retention. This construction obtains collective identifiability through a fixed diagnostic pair; \cref{thm:spectrum} studies scale-dependent task composition.

The sample estimator uses \eqref{eq:sine-moment}, whose signal singular value is bounded away from zero. A sphere-net Bernstein argument bounds its operator-norm error, and singular-vector perturbation gives expected representation error $O(1\wedge k/m)$. The decoder projects the quantized learned function. \Cref{app:nonlinear} gives the arbitrary-decoder reductions, the full concentration argument, and the exact bit accounting.

\paragraph{What has been learned.}
The nonlinear direction $u$ and output loading $a$ are unknown. The link $g$, Gaussian observation interface, and projection-task family are known. Every sample supplies an $r$-dimensional response. The theorem counts observations and retained source-dependent bits separately from real-arithmetic work and public descriptions. It gives an explicit recovery result for the declared sine-ridge family.

\paragraph{Fixed tasks and adaptive stress tests.}
For a coherent learned function $\widehat F$, let $C$ be its conditional output-error second moment. Writing $\Psi_q(C)$ for the sum of its largest $q$ eigenvalues,
\begin{equation}
 \Psi_q(\E C)\le\E\Psi_q(C)
 \le\frac rq\Psi_q(\E C).
 \label{eq:quantifier-factor}
\end{equation}
For $q=r/2$, choosing a stress task after seeing the learned candidate can increase expected loss by at most a factor two. \Cref{thm:nonlinear} and the numerical checks use the stated factor-two comparison between the two orders.

\Needspace{4\baselineskip}
\section{Certifying purposes and accounting for computation}
\label{sec:computation}
The information curve characterizes retention cost. Computational accessibility requires an encoding and decoding procedure. We distinguish a positive certification result from a computational obstruction.

\subsection{Discovering the active purposes}
Consider a compact convex design set $\mathcal C$ in a finite-dimensional Hilbert space and a compact task set $\mathcal G$. Let
\[
f(K)=\sup_{G\in\mathcal G}\langle G,S+K\rangle,
\quad D_{\mathcal C}=\operatorname{diam}(\mathcal C),
\quad L_{\mathcal G}=\sup_G\norm G.
\]
At iteration $t$, find $G_t$ with $f(K_t)\le\langle G_t,S+K_t\rangle+\delta_t$ and update by projected gradient descent against that task. With step size $D_{\mathcal C}/(L_{\mathcal G}\sqrt T)$, the averages $\bar K_T,\bar G_T$ satisfy
\begin{equation}
 L_T:=\min_K\langle\bar G_T,S+K\rangle
 \le f_*\le U_T:=f(\bar K_T),\qquad
 U_T-L_T\le\frac{D_{\mathcal C}L_{\mathcal G}}{\sqrt T}
 +\frac1T\sum_t\delta_t.
 \label{eq:stress-gap}
\end{equation}
The lower certificate can be represented by at most $1+\operatorname{affdim}(\mathcal G)$ returned tasks. This is an application of online convex optimization \cite{zinkevich}. The bound counts $T$ calls returning globally $\delta_t$-accurate tasks and $T$ convex projections; numerical errors in either subproblem must enter the certificate. For projection tasks, task search is a top-eigenspace computation. For unrestricted nonlinear tasks, it may itself be hard. Unlike the Boolean universal-feature construction of Hu et al.\ \cite{hu}, this guarantee starts from a specified convex design set and a task-search procedure.

For an independently learned candidate and noisy validation residuals $R_i=e(X_i)+\xi_i$ with isotropic noise, let $C=\E[ee^\top]$ and $\widehat\Sigma=N^{-1}\sum_iR_iR_i^\top$. The empirical top-$q$ projector $\widehat P_q$ obeys
\begin{equation}
 \Psi_q(C)-\tr(\widehat P_qC)
 \le2q\norm{\widehat\Sigma-(C+\sigma^2I)}_{\op}.
 \label{eq:validation}
\end{equation}
For the bounded sine-ridge model, a sub-Gaussian second-moment bound gives a high-probability certificate. A candidate adapted to the same holdout requires uniform control or fresh validation samples. Proofs are given in \cref{app:stress}.

\subsection{Information equivalence need not be computational equivalence}
\begin{proposition}[A computational boundary]
\label{prop:crypto}
Assume length-preserving one-way permutations exist. Let $X$ be uniform on $\{0,1\}^n$, $n\ge2$, and let future tasks ask for $\langle r,X\rangle\bmod2$. The preparation interfaces $X$ and $f_n(X)$ are Blackwell equivalent and contain the same $n$ bits of information. Their unrestricted zero-loss retention cost is exactly $n$ bits. Under the first interface, linear-time preparation and response suffice. Under the second, every jointly probabilistic polynomial-time encoder--decoder, even retaining the entire observation, has average error at least $1/2-\negl(n)$ for an independent uniform task index $r$.
\end{proposition}

The proof is a direct reduction to the Goldreich--Levin hard-core theorem \cite{goldreichlevin}. Exact answers to the coordinate parities require $n$ retained bits. A polynomial-time decoder predicting a random parity from $f_n(X)$ would contradict hard-core security. \Cref{app:crypto} specifies the randomness and quantifiers.

The following computational separation assumes length-preserving one-way permutations. Natural Gaussian single-index models also exhibit acquisition gaps within statistical-query and low-degree computational models \cite{damian}; those restrictions are part of the lower-bound statement. The first result is conditional on cryptographic hardness; the Gaussian examples use statistical-query and low-degree restrictions. Together they show that computational accessibility depends on the representation interface and the permitted algorithms, in addition to decision value and mutual information.

\section{Consequences and conclusions}
\label{sec:discussion}
A finite system succeeds by preserving the action comparisons that future decisions can use across an environment of unbounded complexity. The joint decision-regret state specifies those comparisons; its information curve gives their retention cost. Weak-star closure determines the general fixed-menu obstruction, while singular mass determines the exact obstruction when all finite menus are admitted together.

Under the compact operational assumptions, the joint law of $X$ is a sufficient description of the entire Bayesian information curve. The counterexamples establish the limits of two simpler summaries: a pairwise supremum metric and unconditional task covariance. The insufficiency results apply to the particular pairwise metric and unconditional covariance summaries analyzed here. Regular diagnostic geometry, the collective Gaussian program, the nested-resolution law, and the exact parity curve give calculable characterizations in distinct structured regimes. Statistical sufficiency characterizes the relevant decision information; acquisition, encoding, and decoding require their own constructions and work bounds.

\begin{table}[t]
\centering\small
\begin{tabularx}{\linewidth}{@{}>{\raggedright\arraybackslash}p{.24\linewidth}>{\raggedright\arraybackslash}X>{\raggedright\arraybackslash}X@{}}
\toprule
Question & Mathematical criterion & What is charged\\
\midrule
Preserving future decision value & $\RR_{\TT}(D)$ and the calibrated decision state & Mutual information; fixed-length rate after iid amortization under compactness\\
Possibility with finite information & Weak-star closure; singular mass for all finite menus & Any finite message from some finite observation prefix\\
Discovering a reusable representation & $\mathcal R_{m,M}$ in a declared observation model & Raw samples and a hard $M$-bit message; worst-parameter risk\\
Finding active future tasks & Primal--dual gap in \eqref{eq:stress-gap} & Task-search and convex-projection calls\\
Using information efficiently & Declared computational model & Encoder and decoder work, including accessible side information\\
\bottomrule
\end{tabularx}
\caption{The theories share the ordering of preparation and use while charging distinct resources. A Bayesian information rate, a minimax finite-sample law, and an algorithmic work bound answer different questions.}
\label{tab:scope}
\end{table}

The collective composition theorem shows that the arrangement of questions can change the high-resolution exponent while leaving pairwise distinctions and individual-question difficulty unchanged. The parity comparison adds a different conclusion: even identical answer covariance can conceal different normalized rate--distortion curves. The joint law of the decision state retains the dependence that these summaries lose.

The nested-resolution theorem provides a constructive answer within a declared hierarchy: retain the source prefix down to the level at which unresolved task cost becomes tolerable. Its block-contraction condition gives an additive-constant guarantee; the polynomial and plateau examples explain why the behavior outside that condition has no single finite-versus-infinite dimension dichotomy.

The main constructive problem is to discover collective structure from raw observations. The nonlinear model learns an unknown direction and loading with a known link, first-moment exposure, and two complementary diagnostic tasks; its direct code has dimension-dependent overhead. The hierarchical model supplies the scale-dependent competition as part of its task description. A decisive extension would infer the relevant hierarchy, source frequencies, and task reach before evaluating held-out combinations, then predict their coding losses without refitting a dimension. Effective rank, task-averaged geometry, and universal-feature complexity provide comparison points. The numerical checks evaluate the specified formulas and constructions. An external test would estimate task structure in a separately trained system and predict held-out coding losses.

Finite retention is governed by the decisions that may be asked together, under the distribution of evidence available before they are asked. Their number, individual difficulty, and pairwise distinctions can all be misleading in isolation. What matters is which distinctions a common representation must preserve for the whole family. This collective requirement is the starting point for explaining how finite information remains useful in an infinite world.

\phantomsection\label{references-start}
\begingroup
\setlength{\bibsep}{2pt}

\endgroup

\clearpage
\phantomsection\label{supplement-start}
\begin{center}
{\Large\bfseries Supplementary proofs}\\[.5em]
{\large Why Finite Intelligence Works in an Infinite World: A Mathematical Theory}
\end{center}
\noindent
This supplement proves the results used in the main argument. All probability spaces admit the stated standard Borel realizations. Null-set identifications are made relative to the fixed experiment. Compact-family arguments use common continuous versions of the task coordinates, which makes the uncountable task suprema measurable and well defined.
\appendix

\section{Canonical states and operational coding}
\label{app:operational}
\subsection{Measurability and calibration}
The space $\mathcal B=C(T\times[d])$ is a separable Banach space when $T$ is compact metric. The compact support $\mathcal E$ is Borel. Its closed convex hull is also compact: for each $\epsilon>0$, approximate $\mathcal E$ by finitely many points; the convex hull of those points is a compact finite-dimensional set and is an $\epsilon$-net for $\conv(\mathcal E)$. Completeness then gives compactness of the closure. A bounded $\mathcal B$-valued random variable is Bochner integrable, so $\E[X\mid U]$ exists and lies in that closed convex hull almost surely.

Evaluation at each $(t,a)$ is continuous linear. Conditional expectation therefore commutes with evaluation. For each task, the additive baseline $B_t(Y)$ is independent of the action and cancels after the conditional minimum. This proves
\[
\Delta_t(U)=\E\min_a\E[X(t,a)\mid U].
\]
The tower property proves calibration of $Z_0=\E[X\mid U]$. Regular conditional probabilities on standard Borel spaces produce a channel $P_{Z_0\mid X}$; regenerating through that channel preserves $(X,Z_0)$ exactly. Every risk in the display above depends only on this joint law. Data processing and the chain rules give \cref{thm:canonical}. No equality of the joint laws of $(Y,Z)$ and $(Y,Z_0)$ is required: the information reduction deliberately removes dependence on evidence outside $X$.

\subsection{Splitting and attainment}
Given a channel $X\to U$, let $Q_U=\mathcal L(X\mid U)$ and $\mu=\mathcal L(Q_U)$. Disintegration gives
\[
\int q\,\mu(dq)=\nu,
\qquad I(X;U)=\int\KL(q\Vert\nu)\,\mu(dq).
\]
The second identity follows from the relative-entropy chain rule, with both sides allowed to be infinite. The task regret is
\[
\E\min_a\E[X(t,a)\mid U]=\int\varphi_t(q)\,\mu(dq).
\]
Conversely, from any $\mu\in\mathcal M(\nu)$ form the joint measure $q(dx)\mu(dq)$. Its $X$ marginal is $\nu$. Disintegrating in the opposite direction gives an encoder from $X$ whose retained state is $q$. This realizes both the information cost and every task regret.

The probability space $\PP(\mathcal E)$ is compact in its weak topology. The barycenter condition is closed, because it is tested against continuous functions on $\mathcal E$. For each $t$, $\varphi_t$ is continuous: it is the minimum of finitely many integrals of continuous bounded functions. The family $\{\varphi_t:t\in T\}$ is uniformly equicontinuous in $t$, since the evaluation functions $x(t,a)$ are uniformly equicontinuous over compact $\mathcal E$. Hence $\mu\mapsto(t\mapsto\int\varphi_t\,d\mu)$ is continuous into $C(T)$.

The function $q\mapsto\KL(q\Vert\nu)$ is lower semicontinuous, for example by its variational representation as a supremum over continuous test functions. Its nonnegative integral is lower semicontinuous in $\mu$. The feasible set in \eqref{eq:splitting} is therefore compact and the cost attains its extended-real minimum. If a feasible split has finite cost, the attained value is finite.

\subsection{Finite approximations without a rate penalty}
We record the approximation step used in achievability. Fix a channel with finite information and task regrets bounded by $D$. First replace it by its calibrated forecast $Z_0$. Partition the compact range of $Z_0$ into finitely many measurable cells of diameter at most $\eta$, and retain their labels $V$. Since $\E[X\mid V]=\E[Z_0\mid V]$, replacing $Z_0$ by $V$ changes each minimum conditional loss by at most $\eta$. Also $I(X;V)\le I(X;U)$.

Next partition $\mathcal E$ into cells of diameter at most $\eta$ and let $S$ be the source-cell label. Replace $P_{V\mid X}$ by its conditional average given $S$. The joint law of $(S,V)$ is unchanged, and the new channel has information $I(S;V)\le I(X;V)$. For any fixed table of task-dependent actions, approximate each source state by a representative of its cell. Each expected loss changes by at most $\eta$ in either experiment. Optimizing the action table cannot increase this comparison error. Thus every task regret for the finite-source channel is at most $D+3\eta$.

Finally, compactness of $\mathcal E\subset C(T\times[d])$ implies that the evaluation profiles $t\mapsto[x\mapsto x(t,a)]$ form a uniformly totally bounded family. Choose tasks $t_1,\ldots,t_L$ so that every task has a representative with
\[
\max_a\sup_{x\in\mathcal E}|x(t,a)-x(t_\ell,a)|\le\eta.
\]
Using the representative's action table costs at most another $\eta$ in regret. We have reduced to a finite source, finite reproduction variable, and finitely many bounded distortion constraints, without increasing the information cost and with uniform error at most $4\eta$.

\subsection{Finite-alphabet simultaneous coding}
For completeness, let $S$ and $V$ have finite alphabets with a prescribed joint law, and let $d_1,\ldots,d_L$ be bounded distortions. Attach to each value of $V$ the finite table of actions used by all $L$ tasks. A codebook of $2^{n(I(S;V)+\epsilon)}$ independent reproduction sequences drawn from $P_V^{\otimes n}$ contains a jointly typical sequence for a typical source sequence with probability tending to one. This follows by the type-counting bounds: the conditional typical set has probability at least $2^{-n(I(S;V)+o(1))}$ under $P_V^{\otimes n}$, so the covering failure probability is bounded by
\[
\exp\{-2^{n(\epsilon-o(1))}\}.
\]
On joint typicality, every one of the finitely many empirical distortions is within $o(1)$ of its prescribed expectation. Off that event, boundedness controls the loss. Averaging over codebooks gives a code with all $L$ expected distortions at most their targets plus $o(1)$; one can bound the sum of the positive deviations to choose a common deterministic code. Standard type approximations handle zero probabilities by restricting to the joint support. This is the usual finite-alphabet rate--distortion achievability argument with finitely many fidelity tests \cite{coverthomas}. Dembo and Weissman \cite[Theorem~1]{dembo2003} give a general-alphabet coding theorem for families of distortion constraints; the finite reduction here uses the task-regret profiles and their uniform compactness.

Apply this lemma after the previous approximations and let $\eta\downarrow0$ sufficiently slowly along a blocklength diagonal. For a general task, decode using its nearest task representative. The uniform comparison proves the achievability direction of \eqref{eq:operational-equality}.

\subsection{Single-letter converse and endpoints}
By conditioning the whole block encoder on $X^n$, any raw-evidence protocol can first be replaced by a channel from $X^n$ preserving all coordinate task regrets. Let $J$ be its message, let $R$ be an independent shared seed, and let $I$ be uniform on $[n]$ and independent. Put
\[
U=(J,R,X_1,\ldots,X_{I-1},I).
\]
Independence of the source coordinates gives
\[
I(X_I;U)=\frac1n\sum_{i=1}^nI(X_i;J,R\mid X_{1:i-1})
=\frac1n I(X^n;J\mid R)
\le\frac1n\log_2|\mathcal J|.
\]
For each fixed task, access to $U$ cannot increase the Bayes risk relative to the actual block decoder, which uses only $(J,R,i,t)$. Thus the induced single-letter state has every task regret at most the corresponding average block regret. It follows that $\RR_{\TT}(D_n)$ is a lower bound on the rate of that code.

The splitting compactness argument implies lower semicontinuity of $D\mapsto\RR_{\TT}(D)$: take near-optimal splits along a convergent sequence of distortion levels, extract a weakly convergent subsequence, and apply the closed constraints and lower semicontinuous cost. Therefore $\limsup D_n\le D$ implies the converse at $D$, including $D=0$. When $\RR_{\TT}(D)=\infty$, finite asymptotic rate is impossible. When it is finite, an attaining or arbitrarily close channel and the diagonal achievability give rates approaching $\RR_{\TT}(D)$ with distortion $D+o(1)$. This proves the stated endpoint convention without assuming exact zero distortion at every finite blocklength.

\subsection{Least-favorable task measures}
Write $F(\mu)=\int c\,d\mu$ and $v_\mu(t)=\int\varphi_t\,d\mu$. The epigraph with slack
\[
\mathcal C=\{(r,f)\in\R\times C(T):
\text{some }\mu\in\mathcal M(\nu)\text{ satisfies }F(\mu)\le r,
\ v_\mu\le f\}
\]
is convex and closed. Closedness follows by compactness of $\mathcal M(\nu)$, continuity of $v_\mu$, and lower semicontinuity of $F$. Separation from a point $(r,D\ind)$ with $r<\RR_{\TT}(D)$ yields a nonnegative scalar multiplier on the cost and a positive continuous linear functional on $C(T)$. Strict feasibility with finite cost forces the cost multiplier to be positive; otherwise a uniformly negative constraint vector contradicts separation. Normalize that multiplier to one.

The Riesz representation theorem identifies the positive functional with a finite nonnegative measure $\Lambda$ on $T$. Weak duality and separation give
\[
\RR_{\TT}(D)=\sup_{\Lambda\ge0}\inf_{\mu\in\mathcal M(\nu)}
\left\{F(\mu)+\int_T(v_\mu(t)-D)\Lambda(dt)\right\}.
\]
Writing $\Lambda=\beta\lambda$, with $\beta=\Lambda(T)$ and $\lambda\in\PP(T)$ when $\beta>0$, gives \eqref{eq:dual}; $\beta=0$ is independent of the choice of $\lambda$. The formula characterizes the supremal dual value; several multipliers may attain that value. A compact closure of original task profiles may contain limits that are not themselves explicitly indexed requests. The measure belongs to that closure, as stated.

\section{The finite-information closure theorem}
\label{app:closure}
Let $\mathcal L_d$ be the weak-star compact cube of $d$ losses taking values in $[0,1]$. For a finite statistic $A$ with labels $1,\ldots,K$, set $k_j(\theta)=\Prob(A=j\mid\Theta=\theta)$. Then
\[
b_A(\ell)=\sum_{j=1}^K\min_{a\le d}
\int\ell_a(\theta)k_j(\theta)P_\Theta(d\theta).
\]
Each $k_j$ is integrable, so this is a continuous functional of $\ell$ in the weak-star topology. Thus its regret supremum over the graph $\mathcal B_{\TT}$ equals its maximum over the closure. Data processing gives $b_A(\ell)\ge b_Y(\ell)$ and therefore the lower bound in \eqref{eq:boundary}.

The function $b_Y$ is upper semicontinuous: it is the infimum over full-evidence decision rules of continuous affine risk functionals. Hence $b_Y(\ell)-c$ attains its maximum $\gamma$ on the compact closure of $\mathcal B_{\TT}$. Fix $r>\gamma$. For each pair $(\ell,c)$ in that closure, retain a Bayes-optimal full-evidence action for $\ell$. This finite statistic has risk exactly $b_Y(\ell)$ for that task. Its regret is below $r$ on a neighborhood of $(\ell,c)$ by continuity. A finite subcover gives finitely many such statistics; retaining their tuple can only improve every Bayes risk. Its uniform regret is below $r$.

If that tuple $A$ depends on all of $Y$, let $A_n$ be its maximum-posterior-probability label given $\HH_n$, with fixed tie breaking. Martingale convergence gives $\Prob(A_n\ne A)\to0$. Using the decoder for $A$ on $A_n$ adds at most this disagreement probability, uniformly over every bounded task and every finite action count. This supplies a finite-prefix code.

Now let $U$ be any representation with joint density
\[
k(\theta,u)=\frac{dP_{\Theta,U}}{d(P_\Theta\otimes P_U)}.
\]
Its Bayes risk is
\[
b_U(\ell)=\int\min_{a\le d}\left[\int\ell_a(\theta)k(\theta,u)P_\Theta(d\theta)\right]P_U(du).
\]
Approximate $k$ in $L^1$ by finite sums of rectangle indicators. The induced risk functional is continuous, and its error is bounded uniformly over all $\ell\in\mathcal L_d$ by the $L^1$ approximation error. Thus $b_U$ is continuous. The same closure lower bound applies. Finite $I(\Theta;U)$ implies domination; finite $I(Y;U)$ implies finite $I(\Theta;U)$ by data processing. Every finite summary has finite $I(Y;A)$. These inclusions give all asserted floor equalities and the rate boundary. The equality concerns infima; attainment at the boundary requires additional conditions.

\section{Singular information and all finite menus}
\label{app:singular}
We prove \cref{thm:singular} with the full evidence available, then use the finite-prefix approximation above.

\begin{lemma}[Finite decisions exposing a singular relation]
\label{lem:small-sets}
For every $\epsilon,\eta>0$, there are finitely many measurable source sets $A_1,\ldots,A_N$, each of prior probability at most $\eta$, and a measurable rule $a(Y)$ such that
\[
\Prob\{\Theta\in A_{a(Y)}\}\ge s-\epsilon.
\]
\end{lemma}
\begin{proof}
Choose Polish topologies generating the standard Borel source and evidence sigma fields. Let $N$ be a Borel singular set with $Q(N)=0$ and $P_{\Theta,Y}(N)=s$. By Fubini, remove a $P_Y$-null set of evidence values so that every vertical section $N_y$ is prior-null. This removal does not change its joint probability. Inner regularity yields a compact $C\subset N$ with joint mass at least $s-\epsilon$.

If $C$ is empty the conclusion is immediate. Otherwise its evidence projection $Y_C$ is compact. In the sum product metric, define
\[
g_\delta(y)=P_\Theta\{\theta:\dist((\theta,y),C)\le\delta\},\qquad y\in Y_C.
\]
The closed-neighborhood indicator is upper semicontinuous in $y$ pointwise in $\theta$; reverse Fatou implies upper semicontinuity of $g_\delta$. As $\delta\downarrow0$, these functions decrease pointwise to $P_\Theta(C_y)=0$. The convergence is uniform on $Y_C$. To see this directly, otherwise some $\delta_n\downarrow0$ and $y_n\to y_*$ satisfy $g_{\delta_n}(y_n)\ge\eta$; for every fixed $\delta>0$, upper semicontinuity would give $g_\delta(y_*)\ge\eta$, contradicting the zero limit.

Choose $\delta$ with $\sup_{Y_C}g_\delta\le\eta$, and a finite $\delta/2$-net $y_1,\ldots,y_N$ of $Y_C$. Put
\[
A_j=\{\theta:\dist((\theta,y_j),C)\le\delta\}.
\]
The prior mass is at most $\eta$. A nearest-center rule, with deterministic tie breaking and arbitrary extension outside $Y_C$, selects a set containing $\theta$ whenever $(\theta,y)\in C$. This proves the stated lower bound.
\end{proof}

\subsection{The lower bound}
For the finite task with loss $\ell_j(\theta)=1-\ind_{A_j}(\theta)$, full evidence has risk at most $1-s+\epsilon$. Suppose $P_{\Theta,U}\ll P_\Theta\otimes P_U$. Absolute continuity gives the modulus
\[
\omega_U(\eta)=\sup\{P_{\Theta,U}(E):(P_\Theta\otimes P_U)(E)\le\eta\}
\longrightarrow0\quad(\eta\downarrow0).
\]
For every decoder $a(U)$, its success event has product probability at most $\eta$, and therefore actual probability at most $\omega_U(\eta)$. Its regret is at least $s-\epsilon-\omega_U(\eta)$. First let $\eta\downarrow0$, then $\epsilon\downarrow0$.

Choose one sequence $\epsilon_j,\eta_j\downarrow0$ and the tasks supplied by Lemma~\ref{lem:small-sets}. These tasks depend only on the original experiment. The absolutely continuous part of full-evidence success tends to zero because the selected event has product probability at most $\eta_j$; the singular part has mass at most $s$. Full-evidence success therefore tends to $s$, while any dominated retained representation has success tending to zero. The regrets tend to $s$ along that same task sequence.

\subsection{The upper bound}
Write $k=dP_{\mathrm{ac}}/dQ$. Approximate $k$ in $L^1(Q)$ by finite rectangular simple functions and refine the finitely many evidence sets they use into a partition $A=A(Y)$. Conditional expectation in the evidence coordinate gives a cellwise constant density $k_A$ with
\[
\norm{k-k_A}_{L^1(Q)}<2\epsilon.
\]
From the finite label $A$, simulate an auxiliary $Y'$ with law $P_Y(\cdot\mid A)$ independently of $\Theta$ given $A$. In the resulting joint law of $(\Theta,Y')$, the absolutely continuous contribution is $k_AQ$. The contribution $S_A$ originating from $P_{\mathrm{s}}$ has mass $s$ and is dominated by $Q$: on each of the finitely many positive-probability cells, its source marginal is dominated by $P_\Theta$ and its simulated evidence law by $P_Y$.

Thus
\[
\TV(P_{\Theta,Y},P_{\Theta,Y'})
\le\tfrac12\norm{k-k_A}_1+\TV(P_{\mathrm{s}},S_A)
\le\epsilon+s.
\]
A decoder using $A$ can simulate $Y'$ and use the full-evidence Bayes rule for any task. Its loss exceeds the full-evidence loss by at most this total variation distance, uniformly over all bounded finite-action losses. Its own Bayes rule is no worse. Hence one finite statistic has worst regret at most $s+\epsilon$. Approximating it by a finite-prefix MAP label adds arbitrarily little uniformly over all tasks. This proves \eqref{eq:singular-law} and the enlarged-class assertions.

\subsection{Interpretations that must remain separate}
Disintegration of \eqref{eq:lebesgue} shows
\[
s=\E\left[\Pi^{\mathrm{s}}_Y(\Theta\text{-space})\right],
\]
where the posterior is decomposed relative to the prior. The obstruction is average posterior mass on prior-null sets. Singular mass measures posterior concentration on prior-null sets, a distinct property from infinite relative entropy.

For $m$ independent copies and \emph{arbitrary whole-block finite-action tasks}, the singular mass is $1-(1-s)^m$. The only absolutely continuous term in the product decomposition is $P_{\mathrm{ac}}^{\otimes m}$, of mass $(1-s)^m$. This whole-block decision class permits an arbitrary finite-action block task; \cref{thm:operational} uses one fixed task and average coordinate loss, leading to a different operational criterion.

\section{Fixed menus and a maximal quantifier separation}
\label{app:compact}
For binary losses, put $f=\ell_1-\ell_2$, so $\norm f_\infty\le1$. Conversely every such $f$ is realized by $(1+f)/2,(1-f)/2$. Cancellation of the common term gives
\begin{equation}
\Delta_2(A)=\frac12\sup_{\norm f_\infty\le1}
\{\norm{Tf}_1-\norm{Q_ATf}_1\},\qquad Q_Ah=\E[h\mid A].
\label{eq:binary-operator}
\end{equation}
Triangle inequality and $L^1\le L^2$ prove the upper bound in \eqref{eq:compact-bounds}.

For $t\in[-1,1]$, apply \eqref{eq:binary-operator} to $(f-t)/2$. Then
\[
\E|Tf-t|-\E|Q_ATf-t|\le4\Delta_2(A).
\]
Integrating $t$ from $-1$ to $1$, using $\int_{-1}^1|x-t|dt=1+x^2$ for $|x|\le1$, and the $L^2$ projection identity, yields
\[
\norm{(I-Q_A)Tf}_2^2\le8\Delta_2(A).
\]
For $d$ actions, write $\ell_a=(1+f_a)/2$. The expected maximum of the $d$ residuals is at most their summed mean-square norm to the power $1/2$. This proves the remaining bound in \eqref{eq:compact-bounds}.

If the binary floor is zero, the finite-rank operators $Q_AT$ approximate $T$ in operator norm, and $T$ is compact. Conversely, let finite evidence partitions generate $\sigma(Y)$. Their conditional expectations converge strongly in $L^2$, and therefore uniformly on the compact closure of the image under $T$ of the $L^\infty$ unit ball. The displayed bounds imply zero regret for every fixed $d$. Finite-prefix implementation follows from the MAP approximation already proved.

The operator in this statement acts from $L^\infty$ to $L^2$. For exact revelation of a countably infinite discrete source, the identity on $L^2$ is not compact, whereas the map from $L^\infty$ to $L^2$ is compact by truncation of a vanishing probability tail. Confusing the two would give a false necessity condition.

\subsection{Independent coordinate revelations}
Let $B_j$ be independent fair bits and let $R_j\sim\Ber(p_j)$ be independent reveal indicators. The observation is $Y_j=(R_j,R_jB_j)$. In the Walsh basis $\chi_S$ indexed by finite subsets $S\subset\N$,
\[
T\chi_S=\chi_S(B)\prod_{j\in S}R_j.
\]
These images are orthogonal and have squared norms $\prod_{j\in S}p_j$. Conditioning on the first $m$ observations removes exactly the terms with $S\not\subseteq[m]$. Consequently
\[
\norm{(I-Q_m)Tf}_2^2
=\sum_{S\not\subseteq[m]}\widehat f(S)^2\prod_{j\in S}p_j
\le\sup_{j>m}p_j\,\norm f_2^2.
\]
For $p_j=1/j$, this tends to zero uniformly over $\norm f_\infty\le1$. The first $m$ observations have at most $3^m$ states, so every fixed menu has zero floor.

Because $\sum_jp_j=\infty$, infinitely many bits are revealed almost surely. Conditional on such evidence, its source consistency set has prior probability zero: matching the first $h$ revealed bits has prior probability $2^{-h}$. The joint law is carried by the consistency relation, but the product law gives it probability zero. Thus $s=1$, and \cref{thm:singular} proves \eqref{eq:quantifier-separation}.

\section{Rates that pairwise metrics cannot determine}
\label{app:metric}
Write $L=2^{d-1}\ge4$. If two source states have the same $B$ and different $J$, a singleton task at one index separates them. If their $B$ values differ and the indices agree, every shared task separates them. If both differ, an index different from both makes the two outputs equal to the two different $B$ values. Hence the shared family separates all pairs, just as coordinate tasks do. Each answer is fair by the independent mask $B$.

For coordinates, a common message with each bit error at most $D\le1/2$ obeys
\[
I(\Theta;U)\ge\sum_{i=1}^d I(B_i;U)
\ge d[1-h_2(D)].
\]
Independent binary symmetric channels attain equality.

For the shared family, any single task is fair and therefore gives the lower bound $1-h_2(D)$. For $D\ge1/L$, retain $U=B\oplus N$ with independent crossover
\[
q=\frac{D-1/L}{1-2/L}\in[0,1/2].
\]
Answer $U$ to every task. Its error is $q+(1-2q)/L=D$ and its information is $1-h_2(q)$. Thus
\begin{equation}
1-h_2(D)\le\RR_{\mathrm{shared}}(D)
\le1-h_2\!\left(\frac{D-1/L}{1-2/L}\right).
\label{eq:shared-bounds}
\end{equation}
At every fixed positive $D<1/2$, the bounds meet as $L\to\infty$. At zero distortion, the complete task-answer pattern identifies the source, forcing $I(\Theta;U)=H(\Theta)=d$ in both families.

\subsection{A limit of covariance descriptions}
\label{app:covariance-limits}
Let $k\ge2$, $N=2^k-1$, and let $W_1,\ldots,W_N$ be independent fair bits. The coordinate answers $(-1)^{W_j}$ have mean zero and covariance $I_N$. The parity answers
\[
 (-1)^{v^\top(W_1,\ldots,W_k)\bmod2},
 \qquad v\in\{0,1\}^k\setminus\{0\},
\]
also have mean zero. Distinct nonzero $v,w$ satisfy $v+w\ne0$ over $\mathbb F_2$, so their product is a nonconstant fair parity and has expectation zero. Their covariance is therefore the same $I_N$.

A retained state answering every coordinate with zero error determines all $N$ independent bits, and hence has information at least $N$. Encoding them attains equality. A retained state answering every listed parity determines the first $k$ bits, since the coordinate parities are included, and those $k$ bits determine every answer. Its exact zero-error rate is therefore $k$. These conclusions also hold for hard finite messages, with the corresponding numbers of states. At common bit error $0\le D\le1/2$, conditional-entropy subadditivity and binary Fano give the coordinate rate $N[1-h_2(D)]$, attained by independent binary test channels in the information-budget model; the parity rate is at most $k$ by exact storage. The source and marginal answer laws agree in the two experiments, but their joint answer laws do not. The example isolates the insufficiency of unconditional task covariance while leaving the pointwise metrics and conditional multiscale structure free to differ.

\subsection{A sharp Fourier--entropy bound and the exact parity converse}
\label{app:parity-retention}
Let $W$ be uniform on $G=\mathbb F_2^k$ and write $m=|G|=2^k$. Every nonzero $v\in G$ indexes the task $v^\top W\bmod2$, with zero--one loss. Define
\[
 R_k^{\rm par}(D)=\inf_{P_{U\mid W}}
 \left\{I(W;U):\sup_{v\ne0}\inf_{\widehat a_v}
 \Prob\{\widehat a_v(U)\ne v^\top W\bmod2\}\le D\right\}.
\]
The task is fixed independently of the source, message, and any independent shared seed, which is included in $U$. The source is fully observed.

For a probability mass function $p$ on $G$, use the Walsh normalization
\[
 \widehat p(v)=\sum_{x\in G}p(x)(-1)^{v^\top x},\qquad
 \beta(p)=\frac1{m-1}\sum_{v\ne0}|\widehat p(v)|.
\]
Define $q_\beta=[1+(m-1)\beta]/m$ and
\[
 H_*(\beta)=h_2(q_\beta)+(1-q_\beta)\log_2(m-1).
\]
The average absolute Fourier coefficient measures the predictability of the entire parity family. The next result bounds the entropy that can coexist with this predictability.

\begin{suppproposition}[Sharp Fourier--entropy envelope]
\label{prop:parity-retention}
For every $k\ge1$ and every probability mass function $p$ on $\mathbb F_2^k$,
\begin{equation}
 H(p)\le H_*(\beta(p)).
 \label{eq:parity-entropy-sharp}
\end{equation}
For each $\beta\in[0,1]$, equality holds for
\[
 p_\beta=\beta\delta_{x_0}+(1-\beta)\Unif(G),\qquad x_0\in G.
\]
Consequently, for every $0\le D\le1/2$,
\begin{equation}
 \boxed{R_k^{\rm par}(D)
 =k-h_2(\varepsilon_k)-\varepsilon_k\log_2(2^k-1),
 \quad\varepsilon_k=2D(1-2^{-k}).}
 \label{eq:parity-retention}
\end{equation}
\end{suppproposition}

We first give the classical moment bound used in the proof. Bennett's bounded-variance inequality and its quadratic-majorant formulation supply the antecedent \cite{bennett1962,seneta1982}. The short argument below includes the variance-upper-bound version and fixes the normalization needed here.

\paragraph{Bounded-variance exponential moments.}
If $T\le1$ almost surely, $\E T=0$, and $\E T^2\le v$ for $v>0$, then
\begin{equation}
 \E e^{tT}\le\frac{v e^t+e^{-vt}}{1+v}\qquad(t\ge0).
 \label{eq:parity-moment}
\end{equation}
Equality is attained by the two-point variable taking values $1$ and $-v$ with probabilities $v/(1+v)$ and $1/(1+v)$.
\begin{proof}
Set $a=-v$ and define the quadratic
\[
 Q_t(x)=e^{ta}\{1+t(x-a)\}
       +d_t(x-a)^2,\qquad
 d_t=\frac{e^t-e^{ta}\{1+t(1-a)\}}{(1-a)^2}\ge0.
\]
It agrees with $e^{tx}$ at $a$ and $1$, and its derivative agrees at $a$. The Hermite interpolation remainder gives, for each $x\le1$,
\[
 e^{tx}-Q_t(x)=\frac{t^3e^{t\xi}}{6}(x-a)^2(x-1)\le0
\]
for some $\xi$ between $x$ and the interpolation nodes. At the nodes or at $t=0$ the conclusion follows directly. Thus $\E e^{tT}\le\E Q_t(T)$. The latter depends only on the mean and second moment; its coefficient of the second moment is $d_t\ge0$. Replacing that moment by $v$ gives the expectation of $Q_t$ under the stated two-point law, where $Q_t$ agrees with the exponential. This proves \eqref{eq:parity-moment}.
\end{proof}

\begin{proof}[Proof of the Fourier--entropy envelope]
Let $Q$ be uniform on $G$. For a fixed $p$, select
$\sigma(v)=\operatorname{sgn}(\widehat p(v))\in[-1,1]$ for $v\ne0$, with zero sign permitted, and put
\[
 f_\sigma(x)=\frac1{m-1}\sum_{v\ne0}\sigma(v)(-1)^{v^\top x}.
\]
Walsh orthogonality gives
\[
 \E_Q f_\sigma=0,\qquad
 \E_Q f_\sigma^2=\frac1{(m-1)^2}\sum_{v\ne0}\sigma(v)^2
 \le\frac1{m-1},\qquad f_\sigma\le1.
\]
Also $\E_p f_\sigma=\beta(p)$. Applying \eqref{eq:parity-moment} with $v=(m-1)^{-1}$ yields
\begin{equation}
 \E_Q e^{t f_\sigma}\le
 M_m(t):=\frac{e^t+(m-1)e^{-t/(m-1)}}m\qquad(t\ge0).
 \label{eq:parity-fourier-mgf}
\end{equation}
For this paragraph only, relative entropy and logarithms are in nats. The variational entropy inequality implies
\[
 \KL(p\Vert Q)\ge t\beta(p)-\ln M_m(t),\qquad t\ge0.
\]
The right side is the convex dual of the two-point law in \eqref{eq:parity-fourier-mgf}. For $0<\beta<1$, its maximum occurs at
\[
 t_\beta=\frac{m-1}{m}\ln\frac{1+(m-1)\beta}{1-\beta},
\]
and equals $\KL(\Ber(q_\beta)\Vert\Ber(1/m))$. At $\beta=0$ take $t=0$; at $\beta=1$ take the limit $t\to\infty$. Since
\[
 \ln m-\KL(\Ber(q)\Vert\Ber(1/m))
 =-q\ln q-(1-q)\ln(1-q)+(1-q)\ln(m-1),
\]
division by $\ln2$ proves \eqref{eq:parity-entropy-sharp}.

For $p_\beta$, every nonzero coefficient is $\beta(-1)^{v^\top x_0}$. Its distinguished atom has mass $q_\beta$, and the other $m-1$ atoms have equal masses $(1-\beta)/m$. Its entropy is exactly $H_*(\beta)$, proving sharpness at every value of $\beta$. No classification of every maximizing distribution is needed.
\end{proof}

\begin{proof}[Proof of the exact parity rate]
For an arbitrary retained state, let $p_u=\mathcal L(W\mid U=u)$. The Bayes error of task $v$ conditional on $u$ is $(1-|\widehat p_u(v)|)/2$. Averaging the constraints over the fixed nonzero parity indices gives
\[
 \E\beta(p_U)\ge1-2D.
\]
The envelope $H_*$ is concave and nonincreasing on $[0,1]$: it is the entropy of the affine family $p_\beta$, with its maximum at the uniform law. Therefore
\begin{equation}
 H(W\mid U)\le\E H_*(\beta(p_U))
 \le H_*(\E\beta(p_U))\le H_*(1-2D).
 \label{eq:parity-posterior-converse}
\end{equation}
It follows that $I(W;U)\ge k-H_*(1-2D)$, which is the right side of \eqref{eq:parity-retention}. Only an average of independently fixed task risks was used; no task is allowed to inspect the realized state or seed.

For attainment, independently draw $T\sim\Ber(2D)$ and $V\sim\Unif(G)$, set $Z=TV$, and retain $U=W\oplus Z$, with neither $T$ nor $V$ available separately. Each nonzero parity of $V$ is fair. Thus $v^\top U$ has error exactly $D$ for every task. The noise has mass $1-\varepsilon_k$ at zero and equal mass $\varepsilon_k/(m-1)$ elsewhere. Since $W$ is uniform, $U$ is uniform, so
\[
 I(W;U)=H(U)-H(U\mid W)
 =k-H(Z)=k-h_2(\varepsilon_k)-\varepsilon_k\log_2(m-1).
\]
This meets the converse, including $k=1$ and both distortion endpoints. For each fixed $k$, the finite-source, finite-task case of \cref{thm:operational} gives the iid amortized coding interpretation.
\end{proof}

On the common $N$-bit source in Appendix~\ref{app:covariance-limits}, the other $N-k$ bits cannot help: averaging an encoder conditionally on its first $k$ source bits preserves every parity risk and does not increase information. Thus \cref{prop:coordinate-parity} follows. The conditional-entropy converse applies to all channels, while the attaining test channel returns a raw $k$-bit vector. The result identifies a mutual-information optimum and iid amortized rate; one-shot alphabet size and efficient iid codebook search are separate problems.

\subsection{The optimal channel and the gain over erasure}
\label{app:parity-refinement}
The preceding converse identifies the optimal channel. We now evaluate its gain over explicit erasure and distinguish information rate from raw output length. Fix $0\le D\le1/2$, let $q=2D$, and take independent $T\sim\Ber(q)$ and $V\sim\Unif(\mathbb F_2^k)$, both independent of $W$. Put $Z=TV$ and retain $U=W+Z$, where addition is over $\mathbb F_2$. The branch indicator $T$ is not retained. Every nonzero parity decoded as $v^\top U$ has error
\[
 \Prob(v^\top Z=1)=q/2=D.
\]
The noise law has one distinguished atom and otherwise equal masses:
\[
 p_Z(0)=1-q+q2^{-k},\qquad
 p_Z(z)=q2^{-k}\quad(z\ne0).
\]
The output $U$ is uniform, so the exact mutual information is $k-H(Z)$. Writing $\epsilon_k=q(1-2^{-k})$ for the nonzero noise mass gives the exact rate
\begin{equation}
 R_k^{\rm par}(D)=B_k(D):=
 k-h_2(\epsilon_k)-\epsilon_k\log_2(2^k-1).
 \label{eq:parity-symmetric-finite}
\end{equation}
Conditioning on the unretained indicator shows $I(W;U,T)=(1-q)k$. Hence $B_k(D)\le(1-q)k$ by data processing, with strict inequality for $0<q<1$: conditional on $U$, the event $W=U$ changes the probability of $T=0$. The sharp converse in Appendix~\ref{app:parity-retention} proves that this gain is optimal in the declared information model.

For fixed $0<q<1$, $\epsilon_k=q+O(2^{-k})$ and $\log_2(2^k-1)=k+O(2^{-k})$. Smoothness of $h_2$ near the fixed $q$ therefore gives
\begin{equation}
 B_k(D)=(1-2D)k-h_2(2D)+O_D(k2^{-k}),
 \label{eq:parity-symmetric-expansion}
\end{equation}
which proves \eqref{eq:parity-symmetric-asymptotic}. The finite formula handles $D=0$ and $D=1/2$ exactly, with values $k$ and zero; the expansion is asserted only at fixed interior distortion. All stated rates use one common channel selected before the task is supplied. Its raw output has $k$ bits; the smaller rate is a mutual-information value and an iid amortized coding limit. The distortion measure is binary zero--one loss, and the construction establishes optimality of the displayed channel among arbitrary encoders.

\section{The collective composition spectrum}
\label{app:spectrum}
We prove \cref{thm:spectrum} first without the mask $B$. The unmasked tasks request $\ind\{J_j\in S\}$ at weight $\alpha_j^2$; denote their values by $V_\beta^0$ and $\RR_\beta^0$.

\subsection{Two finite identities}
For a posterior $p$ on $[N]$ and a uniformly selected $s$-element subset,
\[
\E_Sp_S=\frac sN,\qquad
\E_Sp_S^2=\frac sN\sum_i p_i^2+
\frac{s(s-1)}{N(N-1)}\left(1-\sum_i p_i^2\right).
\]
Subtracting proves \eqref{eq:subset-identity}. For binary probabilities,
\begin{equation}
p(1-p)\le\min(p,1-p)\le2p(1-p).
\label{eq:variance-error}
\end{equation}
Thus the covariance identity bounds binary Bayes error on both sides.

If $J$ is uniform on $[N]$ and $A$ has at most $K$ values, let $q_{ai}=\Prob(A=a\mid J=i)$ and $w_a=N^{-1}\sum_iq_{ai}$. Omitting zero-probability labels gives
\begin{equation}
\E\sum_i\Prob(J=i\mid A)^2
=\sum_a\frac{\sum_iq_{ai}^2}{N^2w_a}
\le\sum_a\frac{\sum_iq_{ai}}{N^2w_a}
\le\frac KN.
\label{eq:collision}
\end{equation}
This allows an arbitrary stochastic encoder of the entire infinite source: condition on the prefix to obtain the displayed channel. The same bound holds conditional on an independent shared seed and then after averaging it.

\subsection{Cardinality converse and construction}
For $N_j=2^j$, the cardinality choice satisfies $N_j^\beta/4\le s_j\le N_j^\beta$ and $s_j\le N_j/2$. Hence
\[
c_j:=\alpha_j^2\frac{s_j(N_j-s_j)}{N_j(N_j-1)}
\ge\frac{a^{-2}}8 N_j^{-\rho_\beta}.
\]
Choose $j$ minimally so that $N_j\ge2K$; then $2K\le N_j<4K$. Averaging task error over the $s_j$-subsets and applying \eqref{eq:collision} gives
\[
\sup_S\Delta_{j,S}(A)\ge c_j(1-K/N_j)
\ge\frac{a^{-2}}{16(4K)^{\rho_\beta}}.
\]
The subset is selected after averaging source and seed randomness; the lower bound therefore holds for an independently fixed task.

For the upper bound, retain the first $n=\lfloor\log_2K\rfloor$ bits. Tasks at depths $j\le n$ are answered exactly. For $j>n$, answer zero; the loss is at most
\[
\alpha_j^2s_j/N_j\le a^{-2}2^{-\rho_\beta j}
\le a^{-2}2^{-\rho_\beta(n+1)}.
\]
This proves $V_\beta^0(K)\asymp K^{-\rho_\beta}$.

\subsection{Information converse across all scales}
Suppose all unmasked task losses after a retained variable $U$ are at most $D$. By \eqref{eq:subset-identity} and \eqref{eq:variance-error},
\[
1-\E\sum_i\Prob(J_j=i\mid U)^2\le8a^2D\,2^{\rho_\beta j}.
\]
The MAP error of the prefix is at most this left side, because $\sum_i p_i^2\le\max_i p_i$. Put $B_*=8a^2$ and
\[
n=\left\lfloor\rho_\beta^{-1}\log_2\frac1{2B_*D}\right\rfloor.
\]
For $k\le n$, bit $X_k$ can be decoded from the length-$k$ prefix estimate with error at most $\tfrac12 2^{-\rho_\beta(n-k)}$. Binary Fano and conditional entropy subadditivity yield
\[
H(X_{1:n}\mid U)
\le\sum_{h=0}^\infty h_2\!\left(\tfrac12 2^{-\rho_\beta h}\right)
=:C_{\rho_\beta}<\infty.
\]
The convergence follows from $h_2(x)\le x\log_2(e/x)$ for $0<x\le1/2$. Therefore
\[
I(X;U)\ge n-C_{\rho_\beta}
\ge\rho_\beta^{-1}\log_2(1/D)-O(1).
\]
The stored-prefix construction uses $\rho_\beta^{-1}\log_2(1/D)+O(1)$ bits and proves the matching upper bound. Using only a single scale in Fano's inequality would not establish this sharp additive $O(1)$ remainder.

\subsection{The common mask and the preserved geometry}
Store $B$ exactly and use any unmasked code to obtain
\[
\RR_\beta(D)\le\RR_\beta^0(D)+1.
\]
Conversely the mask task implies $I(B;U)\ge1-h_2(D)$. Giving $B$ to the decoder can only improve the other decisions and turns their xor labels into membership labels. Since $B$ and $X$ are independent,
\[
I(X;U\mid B)=I(X;U,B)\ge\RR_\beta^0(D).
\]
The chain rule gives the reverse bound
\[
\RR_\beta^0(D)+1-h_2(D)\le\RR_\beta(D).
\]
For hard messages, the upper construction uses one bit for $B$ and the remaining bits for a prefix. The lower bound grants $B$ and conditions \eqref{eq:collision} on $B$ and the independent seed; the message still has at most $K$ values. The $K=1$ case is bounded below by the mask task's error $1/2$. This proves both laws in \cref{thm:spectrum}.

The metric equality \eqref{eq:common-metric} follows because singleton queries exist at every depth. If the first differing source bit is at depth $L+1$, all earlier queries agree and all later responses differ by at most $a^L$; a singleton at depth $L+1$ attains it. The independent mask makes every answer fair, so every individual task at weight $w$ has binary rate $1-h_2(D/w)$ on $0\le D\le w/2$.

At each depth there are finitely many tasks, and their loss amplitudes tend to zero. Adding the null task makes the family compact in the uniform topology. Its regret state is a continuous image of the compact source. Thus \cref{thm:operational} gives the stated amortized interpretation as well as the directly proved hard-cardinality bounds.

\subsection{A source-weighted diagnostic covariance}
The proof extends to finite binary task sets at each nested prefix scale. Write the weighted response vector as $f_t=\alpha_t b_t\in\R^{N_j}$, let $v_t=f_t-\bar f_t\ind$, and put $P_0=I-\ind\ind^\top/N_j$. Define
\begin{align*}
e_j&=\sup_t\alpha_t^2\min\{\Prob(b_t=1),\Prob(b_t=0)\},\\
\gamma_j&=\max_{\lambda\in\PP(T_j)}
\lambda_{\min}\!\left(\E_{t\sim\lambda}v_tv_t^\top\bigm|_{\ind^\perp}\right).
\end{align*}
The minimum eigenvalue in this display is restricted to the zero-sum subspace $\ind^\perp$. For any posterior $p$, the matrix $H(p)=\diag(p)-pp^\top$ is positive semidefinite, annihilates $\ind$, and has trace $1-\norm p_2^2$. Thus a least-favorable task mixture with covariance at least $\gamma_jP_0$ forces average posterior variance at least $\gamma_j(1-\norm p_2^2)$. Equations \eqref{eq:variance-error} and \eqref{eq:collision} give
\[
V(K)\ge\sup_j\gamma_j(1-K/N_j)_+,
\qquad
V(K)\le\sup_{j>\lfloor\log_2K\rfloor}e_j.
\]
If $\gamma_j\ge c2^{-\rho j}$ and $e_j\le C2^{-\rho j}$, the same nested entropy proof gives $V(K)\asymp K^{-\rho}$ and $\RR(D)=\rho^{-1}\log_2(1/D)+O(1)$. The covariance optimization is a finite semidefinite design problem for the specified task family. Its role here is to expose the collective structure that the pairwise supremum discards.

\subsection{General nested resolutions: full proof}
\label{app:nested}
We prove \cref{thm:nested} independently of the special power-law calculation. The level costs and profile need only satisfy \eqref{eq:nested-profile}--\eqref{eq:nested-entropy}. All tasks remain fixed independently of the realized source, message, and common seed.

\paragraph{A useful consequence of entropy summability.}
There is an integer $L\ge1$, depending only on $\mathcal E_c$, such that
\begin{equation}
 c_{j+L}\le\tfrac12c_j\quad\text{for every }j\ge1.
 \label{eq:nested-block-decay}
\end{equation}
Choose $L$ with $(L+1)h_2(1/4)>\mathcal E_c$. If $c_{j+L}>c_j/2$, monotonicity gives $c_{j+L}/(2c_i)>1/4$ for every $j\le i\le j+L$. These $L+1$ terms alone contradict \eqref{eq:nested-entropy}. Consequently, for every fixed $u>0$,
\begin{equation}
 |j(uD)-j(D)|=O_u(1),\qquad
 |\log_2N_{j(uD)}-\log_2N_{j(D)}|=O_u(1).
 \label{eq:nested-crossing}
\end{equation}
For $u\ge1$, iteration of \eqref{eq:nested-block-decay} gives
$j(D)\le j(uD)+L\lceil\log_2u\rceil$; the other case follows by reversing the comparison. The second bound uses $N_{j+1}/N_j=b_{j+1}\le b_*$. This is the missing uniform control when a constant multiplying distortion is absorbed into an additive rate remainder.

\paragraph{Hard-message upper bound.}
Let $n=n(K)$. The message $(B,J_n)$ has exactly $2N_n\le K$ possible values. It answers every task at level $j\le n$ exactly. For $j>n$, output $B$, whose weighted error is at most $w_js_j/N_j\le a_+c_j\le a_+c_{n+1}$. The mask task is exact. This proves the upper bound in \eqref{eq:nested-cardinality}. A binary description uses $\lceil\log_2(2N_n)\rceil$ bits; no decoder-accessible mask is supplied for free.

\paragraph{Hard-message converse.}
Let the retained state be $(A,R)$, where $R$ is an independent shared seed and $A$ has at most $K$ values conditional on each seed. Giving $B$ to the decoder can only decrease every task error. Conditional on each $(B,R)$, $J_j$ is uniform on its $N_j$ regions. Thus \eqref{eq:collision} applies to the conditional stochastic channel, even if the encoder uses the whole infinite source. For a subset $S$ uniform among the $s_j$-subsets, \eqref{eq:subset-identity} and \eqref{eq:variance-error} imply
\begin{align}
 \sup_{S:\,|S|\le s_j}\Delta_{j,S}(A,R)
 &\ge \frac{w_js_j(N_j-s_j)}{N_j(N_j-1)}
 \left(1-\frac K{N_j}\right)_+\nonumber\\
 &\ge \frac{a_-c_j}{2}\left(1-\frac K{N_j}\right)_+.
 \label{eq:nested-level-lower}
\end{align}
The subset is averaged independently of $(B,R)$; a fixed task chosen before the message attains at least that average.

Choose $j_*$ minimally with $N_{j_*}\ge2K$. If $n=n(K)$, then $j_*\ge n+1$. Since $N_{n+1}>K/2$ and every refinement has at least two children, $N_{n+3}>2K$, so $j_*\le n+3$. Equation \eqref{eq:nested-profile} gives $c_{j_*}\ge\eta^2c_{n+1}$. Substitution in \eqref{eq:nested-level-lower} proves the lower bound in \eqref{eq:nested-cardinality}. The argument for the cardinality lower bound uses adjacent-ratio control alone.

\paragraph{Information converse across all refinements.}
Let an arbitrary retained variable $U$ satisfy every task-error constraint at level $D$. Include any independent shared seed in $U$. Reveal $B$ for analysis and write
$p^{(j)}_i=\Prob(J_j=i\mid U,B)$.
The same subset calculation gives
\begin{equation}
 1-\E\norm{p^{(j)}}_2^2\le\frac{2D}{a_-c_j}.
 \label{eq:nested-collision-information}
\end{equation}
The MAP error for $J_j$ is at most this left side, since $\sum_i p_i^2\le\max_i p_i$. A decoded prefix supplies a decoded refinement digit $W_j$, so the same error bound holds for that digit.

Put $n_-=j(4D/a_-)-1$ and take $D$ small enough that $n_-\ge1$. For $i\le n_-$, define
\[
 q_i=\frac{c_{n_-}}{2c_i}\le\frac12.
\]
Since $c_{n_-}>4D/a_-$, the digit error is bounded by $q_i$. Finite-alphabet Fano and monotonicity of its bound on $[0,1/2]$ yield
\[
 H(W_i\mid U,B)\le h_2(q_i)+q_i\log_2(b_i-1).
\]
Use $h_2(q)\ge2q$ on $[0,1/2]$ and conditional-entropy subadditivity to obtain
\begin{equation}
 H(J_{n_-}\mid U,B)\le C_b\mathcal E_c,
 \qquad C_b=1+\tfrac12\log_2(b_*-1).
 \label{eq:nested-fano-tail}
\end{equation}
The uncompressed prefix is uniform and independent of $B$, hence $H(J_{n_-}\mid B)=\log_2N_{n_-}$. The separate mask task gives $I(B;U)\ge1-h_2(D)$ for $D\le1/2$. Therefore
\begin{equation}
 I(B,W_1,W_2,\ldots;U)
 \ge 1-h_2(D)+\log_2N_{n_-}-C_b\mathcal E_c.
 \label{eq:nested-information-lower}
\end{equation}
The bounded refinement alphabet permits the digitwise entropy bound required for the additive-constant converse.

\paragraph{Information construction and the sharp additive remainder.}
Let $n_+=j(D/a_+)-1$ and retain $(B,J_{n_+})$. Its loss is at most $D$ and its information cost is $1+\log_2N_{n_+}$. Thus
\begin{equation}
 1-h_2(D)+\log_2N_{n_-}-C_b\mathcal E_c
 \le\RR_{\TT_{\rm nest}}(D)
 \le1+\log_2N_{n_+}.
 \label{eq:nested-finite-sandwich}
\end{equation}
Equation \eqref{eq:nested-crossing} and bounded branching compare both prefix lengths to $j(D)$ within $O(1)$ bits. This proves \eqref{eq:nested-rate}. Combining it with the explicit prefix code shows that its bit length is within an additive constant of the optimal information cost as $D\downarrow0$.

\paragraph{Operational meaning.}
The source is a compact product of finite spaces. Each task depends on finitely many coordinates, there are finitely many tasks at each level, and $w_j\to0$. Adding the null task makes the loss-profile family compact in the uniform topology. The map from the source into the joint decision-regret state is continuous: fixing a sufficiently long prefix fixes every coarse coordinate, while all remaining coordinates have uniformly small amplitude. Its range is compact. The hypotheses of \cref{thm:operational} therefore hold with two actions, giving the stated iid fixed-length interpretation. The direct hard-cardinality and information proofs above do not invoke that operational theorem or identify its budget with a single message length.

\paragraph{Recovering the spectrum and obtaining a non-power example.}
For the original hierarchy, $N_j=2^j$ and
\[
 \frac{a^{-2}}4\,2^{-\rho_\beta j}
 \le\frac{\alpha_j^2s_j}{N_j}
 \le a^{-2}2^{-\rho_\beta j}.
\]
Choose $c_j=2^{-\rho_\beta j}$. Its entropy tail is bounded by $\sum_{h\ge0}h_2(2^{-1-\rho_\beta h})<\infty$, and the theorem recovers \eqref{eq:spectrum-rate}. Allowing a comparable profile, rather than requiring the exact costs to be monotone, is useful because floor operations in $s_j$ can create small upward jumps.

For \eqref{eq:nested-loglog}, the exact level cost is
$c_j=2^{-1-\rho j}(j+1)^{-\zeta}$.
It has adjacent ratio at least $2^{-\rho-\zeta}$, and its backward entropy sum is no greater than the preceding geometric sum with exponent $\rho$. If $L_D=\log_2(1/D)$, the crossing index satisfies
\[
 \rho j(D)+\zeta\log_2(j(D)+1)=L_D+O(1).
\]
First this gives $j(D)\asymp L_D$; substituting back gives
$j(D)=\rho^{-1}L_D-(\zeta/\rho)\log_2L_D+O(1)$.
This proves the displayed correction without fitting an effective dimension.

\subsection{Block contraction and rates beyond the summable regime}
\label{app:decay-regimes}
The entropy-tail condition of \cref{thm:nested} describes a uniform decrease across refinement levels. It is useful to identify that condition precisely before examining profiles outside it.

\begin{suppproposition}[Entropy summability and block contraction]
\label{prop:block-equivalence}
For a nonincreasing positive sequence $c_j$, let
\[
 \mathcal E_c=\sup_{n\ge1}\sum_{j=1}^n
 h_2\!\left(\frac{c_n}{2c_j}\right).
\]
Then $\mathcal E_c<\infty$ if and only if some integer $L\ge1$ satisfies $c_{j+L}\le c_j/2$ for every $j\ge1$. Monotonicity of $c_j$ suffices for this equivalence.
\end{suppproposition}
\begin{proof}
The forward implication is the argument for \eqref{eq:nested-block-decay}: choose $L$ with $(L+1)h_2(1/4)>\mathcal E_c$. Failure of halving on one interval $j,\ldots,j+L$ makes all $L+1$ corresponding entropy terms exceed $h_2(1/4)$.

Conversely, iterating fixed-block halving and using monotonicity gives
\[
 \frac{c_n}{c_j}\le2^{-\lfloor(n-j)/L\rfloor}\quad(1\le j\le n).
\]
Binary entropy is increasing on $[0,1/2]$, hence
\[
 \sum_{j=1}^n h_2\!\left(\frac{c_n}{2c_j}\right)
 \le L\sum_{a=0}^\infty h_2(2^{-a-1})<\infty.
\]
The last series converges by $h_2(x)\le x\log_2(e/x)$ for $0<x\le1/2$.
\end{proof}

\begin{suppproposition}[Two nonsummable refinement profiles]
\label{prop:nonsummable-profiles}
Let the source consist of independent fair binary refinement digits and a fair mask bit $B$. At depth $j$, allow all subset tasks of reach at most $s_j=N_j/2$, where $N_j=2^j$, and give them weight $w_j$. Include the unit-weight mask task. Then:
\begin{enumerate}[label=(\roman*),leftmargin=*]
 \item If $w_j=j^{-p}$ for fixed $p>0$, then $c_j=\tfrac12j^{-p}$, $\mathcal E_c=\infty$, and $\RR(D)=\Theta(D^{-1/p})$ as $D\downarrow0$.
 \item There are nonincreasing weights $w_j\to0$ with $c_{j+1}/c_j\in\{1,1/2\}$ and $\mathcal E_c=\infty$, but $\RR(D)=\Theta(\log_2(1/D))$.
\end{enumerate}
The rates use a single retained channel for all independently selected future tasks. They also have the iid operational interpretation.
\end{suppproposition}
\begin{proof}
For either profile, storing $(B,J_n)$ costs $n+1$ bits and incurs worst loss at most $c_{n+1}$. The subset argument in \eqref{eq:nested-collision-information} uses the level costs and the subset collision identity directly. Here the level cost equals $c_j$, so after revealing $B$ for the converse, the prefix MAP error obeys $\delta_j\le2D/c_j$. If $c_j\ge8D$, then $\delta_j\le1/4$. Fano for the uniform $2^j$-valued prefix yields
\begin{equation}
 I(B,W_1,W_2,\ldots;U)\ge I(J_j;U\mid B)
 \ge\tfrac34j-1.
 \label{eq:nonsummable-fano}
\end{equation}
This lower bound allows arbitrary retained channels and independent common randomness.

For (i), choose $n=\lceil(2D)^{-1/p}\rceil$ for the upper code, and $j=\lfloor(16D)^{-1/p}\rfloor$ for the converse, for sufficiently small $D$. Both are of order $D^{-1/p}$. For $\lceil n/2\rceil\le j\le n$,
$c_n/(2c_j)=(j/n)^p/2\ge2^{-p-1}$, so the partial entropy sum grows linearly in $n$ and $\mathcal E_c=\infty$.

For (ii), define an integer sequence $r_1=1$. At stage $m=1,2,\ldots$, append $m$ transitions with $r$ unchanged followed by $m$ transitions that increase $r$ by one. Put $c_j=2^{-r_j}$ and $w_j=2c_j$. Then $0<w_j\le1$, $w_j\to0$, and every adjacent cost ratio is one or one half. To check the envelope explicitly, at the beginning of stage $m$ the index and exponent are
\[
 j_m=1+m(m-1),\qquad a_m=1+\frac{m(m-1)}2.
\]
For $0\le t\le2m$ transitions into that stage,
$r_{j_m+t}=a_m+\max(0,t-m)$.
The expression $4r_{j_m+t}-(j_m+t)$ is minimized at $t=m$ and equals $(m-1)^2+2>0$. Also $r_j\le j$, because the exponent increases by at most one per transition. Hence
\[
 j/4\le r_j\le j,\qquad 2^{-j}\le c_j\le2^{-j/4}.
\]
The flat portions have unbounded length. At the end of a flat portion of length $m$, at least $m+1$ entropy terms equal $h_2(1/2)=1$, forcing $\mathcal E_c=\infty$.

A prefix of depth $\lceil4\log_2(1/D)\rceil$ gives the logarithmic upper bound. Taking $j=\lfloor\log_2(1/(8D))\rfloor$ gives $c_j\ge2^{-j}\ge8D$ and the matching logarithmic lower bound from \eqref{eq:nonsummable-fano}.

Both constructions have finitely many tasks at each level and $w_j\to0$. The same compactness argument as in Appendix~\ref{app:nested} applies, giving the operational interpretation without invoking the entropy-tail conclusion of \cref{thm:nested}.
\end{proof}

The polynomial profile has $\RR(D)/\log_2(1/D)\to\infty$, while the established bounds for the flat-interval profile give logarithmic rate order. Both profiles violate entropy summability, yielding distinct information-dimension behavior. The equivalence in \eqref{eq:block-equivalence} characterizes the regularity hypothesis of the nested-resolution theorem.

\section{Geometric rates, forgetting, and reuse}
\label{app:reuse}
\begin{proposition}[A sufficient regular geometry for a $k/2$ coefficient]
\label{thm:geometric}
Let a fully observed parameter $Z$ have a density on a compact set $\mathcal Z\subset\R^k$ with $h(Z)>-\infty$. Consider deterministic Hilbert-valued responses $q_{z,t}$ and squared prediction loss. Suppose $\mathcal Z$ has covering number at most $(C/\epsilon)^k$, all response maps are uniformly Lipschitz, and a finite tuple of tasks satisfies
\[
\sum_{\ell=1}^L\norm{q_{z,t_\ell}-q_{z',t_\ell}}^2
\ge c\norm{z-z'}^2\quad\text{for all }z,z'\in\mathcal Z.
\]
Then the squared-prediction information function obeys
\[
\RR^{\mathrm{sq}}_{\TT}(D)=\frac k2\log_2\frac1D+O(1).
\]
The remainder is for fixed $k$, source, and geometric constants.
\end{proposition}
\begin{proof}
A covering code of radius $\sqrt{D}/L_0$, where $L_0$ is the common Lipschitz bound, costs at most $(k/2)\log_2(1/D)+O(1)$ bits and controls every task. For the converse, evaluate the $L$ diagnostic decoders on the same retained variable and project their response tuple onto an $\eta$-nearest feasible tuple. The true tuple is feasible, so the inferred parameter has expected squared error at most $(4L/c)D+o(1)$. Differential entropy and maximum entropy at fixed mean-square error give
\[
I(Z;U)\ge h(Z)-\frac k2\log_2\left(\frac{2\pi e}{k}\E\norm{Z-\widehat Z}^2\right).
\]
Let $\eta\downarrow0$. This proves the result. The argument directly bounds the prediction rate for the stated Hilbert-valued model; the finite-action coding theorem retains its separate hypotheses.
\end{proof}

\subsection{Independent direct sums}
For independent states, the chain rule and independence give
\[
I(X_1,\ldots,X_m;U)=\sum_i I(X_i;U\mid X_{<i})
\ge\sum_iI(X_i;U).
\]
Each induced marginal channel is admissible for its own task problem, so $I(X_i;U)\ge\RR_i(D_i)$. Independent near-optimal marginal channels attain the sum, proving \eqref{eq:direct-sum}.

For an additive product task, optimization over the tuple of actions separates, even when the states are dependent. The largest task regret is $\sum_iw_i D_i(U)$ because tasks can be chosen componentwise. Taking the infimum over admissible error allocations proves \eqref{eq:infconv} in the independent case.

\subsection{Dependent states}
The relative-entropy chain rule gives \eqref{eq:tc-identity}; for finite alphabets it follows immediately by expanding the four entropies in each mutual-information term. Finite-alphabet approximation extends it whenever the displayed divergences are finite. Nonnegativity of $\TC(X\mid U)$ gives
\[
I(X;U)\ge\sum_i\RR_i(D_i(U))-\TC(X).
\]
For the upper bound, use independently randomized marginal channels conditional on the source vector. Their total mutual information is at most the sum of their marginal information costs: the output entropy is subadditive and the conditional output law factors. The compact operational approximation permits finite-output channels arbitrarily close to optimal, avoiding infinite-entropy subtraction. This proves \eqref{eq:reuse-bound}.

If $\RR_i(D_i)=(k_i/2)\log_2(1/D_i)+O(1)$ near zero, Lagrange minimization under $\sum_iw_iD_i\le D$ allocates $D_i=k_iD/(w_i\sum_jk_j)$ to leading order. The sum rate has coefficient $\sum_i k_i/2$. A finite total-correlation subtraction leaves it unchanged. These statements assume a fixed finite number of components and fixed regularity constants.

\section{Exact binary acquisition and retention}\label{app:binary}
\label{app:fr-tools}
We first reduce arbitrary bounded binary decisions and arbitrary encoders to a scalar geometric problem. The discrete acquisition experiment is then compared uniformly with an affine posterior, at an error smaller than the joint asymptotic scale.

\subsection{All bounded decisions reduce to constrained curvature}
For a binary source and posterior $p$, a finite decision problem has Bayes envelope
\[
 g(p)=\min_a\{(1-p)\ell(0,a)+p\ell(1,a)\}.
\]
It is concave and piecewise linear. Its curvature measure $\mu\ge0$ satisfies
\begin{equation}
 g(p)=\text{affine term}-\int_0^1(p-t)_+\,\mu(dt),\quad
 \int t\,\mu(dt)\le1,\quad\int(1-t)\,\mu(dt)\le1.
 \label{eq:fr-curvature}
\end{equation}
Indeed, the intercept at zero of the active supporting line increases by $t\,d\mu(t)$ at a change of slope, while its value at one decreases by $(1-t)\,d\mu(t)$. Both values remain in $[0,1]$. In particular $\mu([0,1])\le2$.

Conversely, every finitely supported nonnegative measure satisfying the two inequalities is realized by a bounded finite decision. Use
\[
 g(p)=\int\min\{(1-t)p,t(1-p)\}\,\mu(dt).
\]
Its finitely many supporting lines have endpoint values in $[0,1]$ and can be taken as the action losses. Any finite measure satisfying the constraints can be approximated for this purpose by finite-action envelopes: all supporting slopes have magnitude at most one, so a dense finite selection of supporting lines converges uniformly. Regret changes by at most twice that uniform approximation error. Thus continuous curvature measures can supply lower bounds for the supremum over finite decisions, without declaring an infinite action set admissible.

If $P$ is the full-evidence posterior and $M$ the retained posterior, both have the same mean and
\begin{equation}
 e(t)=\E(P-t)_+-\E(M-t)_+\ge0.
 \label{eq:fr-stoploss}
\end{equation}
For a task with curvature $\mu$, regret is $\int e\,d\mu$. Linear-program duality gives
\begin{equation}
 \mathcal U(e)=\inf_{x,y\ge0}\{x+y:e(t)\le xt+y(1-t)\text{ for all }t\in[0,1]\}.
 \label{eq:fr-envelope}
\end{equation}
For the finite posterior laws used below this is an ordinary finite linear program. For a general continuous $e$, approximate the interval by nested finite grids. Uniform continuity makes their constraints dense, and compactness of a bounded feasible dual set proves convergence to \eqref{eq:fr-envelope}. The curvature realization above gives the matching decision supremum.

\subsection{Why arbitrary stochastic encoders cannot improve the interval optimum}
Let $Q$ be any posterior with a law on $[0,1]$. Represent it as $Q=F_Q^{-1}(W)$ for $W\sim\Unif[0,1]$, randomizing uniformly inside the appropriate probability interval at an atom. Order the positive-probability encoder messages by their posterior means, and partition the $W$ interval into consecutive cells of those message probabilities in the same order.

For any threshold $t$, the old messages whose means exceed $t$ have total mass $s$. Of all randomized selections of mass $s$ from the posterior law, the upper posterior tail maximizes the integral of $q-t$. The new interval code has a union of upper cells equal to that tail. Its optimal threshold action performs at least as well as choosing this particular union. Therefore the new retained mean has at least as large a stop-loss expectation at every $t$, and the same expectation as $Q$. Equation~\eqref{eq:fr-curvature} shows that it improves every later bounded finite decision simultaneously.

This argument includes atoms and private randomized encoders. If the original observation contains additional variables conditionally independent of the binary target given $Q$, they only randomize a channel out of $Q$ and are already covered. With a shared seed the argument is applied separately for each seed value. It never gives the private randomization tape to the decoder.

\subsection{Exact finite-code evaluation}
Suppose the full posterior has density $f$ and an interval cell is $[a,b]$, with conditional mean $m$. Only the cell containing $t$ contributes to the stop-loss gap. Direct calculation gives
\begin{equation}
 e_C(t)=\min\left\{\int_a^t(t-p)f(p)\,dp,
                          \int_t^b(p-t)f(p)\,dp\right\}.
 \label{eq:fr-cell}
\end{equation}
The first branch applies for $t\le m$, the second for $t\ge m$. They coincide at $m$. Each branch is convex, and $e_C$ vanishes at the boundaries. Hence an affine majorant on $[0,1]$ is valid exactly when it is valid at every cell mean and at the endpoints. Since $x,y\ge0$ already impose the endpoint conditions,
\begin{equation}
 \mathcal U_f(C)=\min_{x,y\ge0}
 \left\{x+y:xm_i+y(1-m_i)\ge e_C(m_i)\text{ for every cell }i\right\}.
 \label{eq:fr-code-lp}
\end{equation}
For a finite mixture of partitions, the averaged profile is convex between successive retained posterior means from all partitions: there its second derivative is $f$. The same finite reduction applies to their union of means. The dual gives a finite downstream task certifying the code's computed worst-task regret. This optimizes the decoder's downstream decision, not the choice of partition.

\subsection{The joint limit}
\label{app:fr-joint}
The proof first removes the discrete posterior grid with a uniform error bound. It then matches a decision-theoretic converse with a smooth one-dimensional code. This order controls both the observation limit and the increasing number of retained states.

\subsection{The exact posterior and its affine comparison}
Integration of the binomial likelihood against a uniform prior gives $\Prob(S=s)=1/(n+1)$ and $\E[P\mid S=s]=(s+1)/(n+2)$. Conditional on $S$, the order of the Bernoulli observations carries no further information about $P$ or $\Theta$. The interval reduction above therefore covers every encoder of the full observation vector.

Put
\[
 m=n+1,\quad a_n=\frac1{2(n+2)},\quad c_n=\frac{n+1}{n+2},\qquad
 \widetilde v_n(w)=a_n+c_nw.
\]
The step posterior $v_n$ in \eqref{eq:fr-step} equals midpoint quantization of $\widetilde v_n$ on the $m$ equal probability intervals. Define
\[
 D_n(w)=\int_0^w[v_n(z)-\widetilde v_n(z)]\,dz.
\]
On every such interval it is a nonnegative concave quadratic, zero at both endpoints, with
\begin{equation}
 0\le D_n(w)\le\frac1{8(n+1)(n+2)}=:d_n.
 \label{eq:fr-primitive}
\end{equation}
For a retained cell of probability width $h_i$, the two posterior means differ by $[D_n(b_i)-D_n(a_i)]/h_i$. Consequently, for every partition into at most $K$ cells,
\[
 \sum_i h_i|m_i-\widetilde m_i|\le Kd_n.
\]
The hinge is one-Lipschitz. Thus the stop-loss profiles relative to the common full-posterior baseline $P\sim\Unif[0,1]$ differ uniformly by at most $Kd_n$, and their universal regrets differ by at most $2Kd_n$. This remains true after averaging any shared seed. Denoting the affine-posterior optimization by $\widetilde V$, we have
\begin{equation}
 |V_{n,K}^{\star}-\widetilde V_{n,K}^{\star}|
 \le\frac{K}{4(n+1)(n+2)},\qquad\star\in\{\mathrm{priv},\mathrm{pub}\}.
 \label{eq:fr-surrogate}
\end{equation}
The interval reduction makes the comparison two-sided: both optimization problems reduce to the same class of quantile interval partitions. At $K\asymp\sqrt n$, its error is $o(n^{-1})$.

The full-posterior baseline remains uniform, while $\widetilde v_n(W)$ is uniform on $[a_n,1-a_n]$. Its acquisition stop-loss gap is
\begin{equation}
 e_n^{\rm acq}(t)=\frac{t(1-t)}{2(n+1)}-
                      \frac1{8(n+1)(n+2)}
 \quad(a_n\le t\le1-a_n).
 \label{eq:fr-acqgap}
\end{equation}
For $t<a_n$ it is $t^2/2$, with the reflected formula above $1-a_n$. Therefore
\begin{equation}
 \sup_{t\in[0,1]}|ne_n^{\rm acq}(t)-b(t)|\longrightarrow0,
 \qquad b(t)=t(1-t)/2.
 \label{eq:fr-baselimit}
\end{equation}
Compression of the affine posterior adds an ordinary nonnegative interval-code gap. At the mean of a probability cell of width $h$, that gap is exactly $c_nh^2/8$.

\subsection{Private retention: a converse without assuming a symmetric code}
Equal-width cells already give a regret of order $n^{-1}$ for $K\asymp\sqrt n$. Consider any sequence within $o(n^{-1})$ of the private optimum. Pass to a subsequence on which its scaled regret tends to $C<\infty$. Its maximum cell width is $O(n^{-1/2})$: at each cell mean the nonnegative compression contribution is $c_nh^2/8$, and the universal regret is at least that value.

Write $h_\rho(t)=\rho t+(1-\rho)(1-t)$. For each code choose the optimal affine majorant in \eqref{eq:fr-envelope}, write it as $U_nh_{\rho_n}$, and pass to a further subsequence $\rho_n\to\rho$. Because the total profile dominates $e_n^{\rm acq}$, its limit satisfies
\[
 s(t):=Ch_\rho(t)-b(t)\ge0\quad\text{on }[0,1].
\]
At each cell mean $t_i$, with probability width $h_i$,
\[
 \frac{nc_nh_i^2}{8}\le s(t_i)+o(1),
\]
where the remainder is uniform. Fix $\epsilon>0$. For all sufficiently large $n$,
\[
 \sum_i\frac{h_i}{\sqrt{s(t_i)+\epsilon}}
 \le K\sqrt{\frac8{nc_n}}.
\]
The mesh vanishes and $t_i$ differs from the probability midpoint by $O(n^{-1})$. Riemann convergence gives
\begin{equation}
 \int_0^1\frac{dt}{\sqrt{Ch_\rho(t)-b(t)+\epsilon}}
 \le\sqrt8\,\kappa.
 \label{eq:fr-private-integral}
\end{equation}
No symmetry of the original code has been assumed. Reflection $t\mapsto1-t$ exchanges $\rho$ and $1-\rho$. Convexity of $x\mapsto x^{-1/2}$ shows that the left side is at least its value at $\rho=1/2$. Sending $\epsilon\downarrow0$ yields
\begin{equation}
 \int_0^1\frac{dt}{\sqrt{C-t(1-t)}}\le2\kappa.
 \label{eq:fr-cothintegral}
\end{equation}
We have $C\ge1/4$ by evaluating the majorant at $t=1/2$. Equality would make the integral infinite. For $C>1/4$ the integral equals
\[
 2\operatorname{arsinh}\!\left(\frac1{2\sqrt{C-1/4}}\right).
\]
Solving \eqref{eq:fr-cothintegral} gives $C\ge\coth^2(\kappa)/4$.

For attainment put $C=\coth^2(\kappa)/4$ and use point density
\begin{equation}
 \lambda_\kappa(t)=\frac1{2\kappa\sqrt{C-t(1-t)}}.
 \label{eq:fr-jointdensity}
\end{equation}
Its integral is one. Its inverse cumulative distribution is \eqref{eq:fr-compander} with $a=\kappa$. On equal-mass cells the compression profile is at most
$[8K^2\lambda_\kappa(t)^2]^{-1}+o(K^{-2})$, uniformly in $t$. This follows either from the exact affine cell formula or the bounded derivative of the inverse compander. Since
\[
 b(t)+\frac1{8\kappa^2\lambda_\kappa(t)^2}=C/2,
\]
the total stop-loss profile is bounded above by $C/(2n)+o(n^{-1})$. Every admissible curvature measure has mass at most two, so regret is at most $C/n+o(n^{-1})$. Using $a=K/\sqrt n$ rather than its limit changes none of these uniform estimates. Equation~\eqref{eq:fr-surrogate} transfers the conclusion to the exact discrete posterior.

\subsection{Weighted interval quantization}
We state the integral estimate used for the shared converse. Suppose a posterior density $f$ and a weight $\psi$ are continuous and strictly positive on a compact interval. Let $e_C$ be its interval-encoding stop-loss gap. Then
\begin{equation}
 \inf_{|C|-1\le K}\int\psi(t)e_C(t)\,dt
 \sim\frac1{24K^2}\left[\int(f(t)\psi(t))^{1/3}\,dt\right]^3.
 \label{eq:fr-weighted}
\end{equation}
Here $f$ is the density in the ordinary coordinate, including its normalization on the interval.

For a cell of width $h$, \eqref{eq:fr-cell} and uniform continuity give its contribution as $f(t_i)\psi(t_i)h^3/24\,[1+o(1)]$, uniformly over cells whose maximum width tends to zero. This also follows by bounding the cell above and below using its smallest and largest densities and weights. An $O(K^{-2})$ comparison code exists. Any nearly optimal code consequently has maximum width $O(K^{-2/3})$, from the positive lower bounds on $f$ and $\psi$, so the uniform approximation applies. H\"older's inequality gives
\[
 \sum_i f(t_i)\psi(t_i)h_i^3
 \ge\frac1{K^2}\left[\sum_i(f(t_i)\psi(t_i))^{1/3}h_i\right]^3.
\]
Riemann convergence proves the converse. Equal-mass cells under density proportional to $(f\psi)^{1/3}$ prove attainment. The proof is uniform for densities converging uniformly to a positive density and intervals with convergent endpoints. In particular it applies to the affine posterior on $[a_n,1-a_n]$ as $n\to\infty$.

\subsection{Shared retention: a fixed least-favorable future decision}
Put $a=\sqrt3\kappa$, $C=\coth^2(a)/4$, and choose
\[
 \psi(t)=\frac{2[C-t(1-t)]^{-3/2}}
                 {\int_0^1[C-z(1-z)]^{-3/2}\,dz}.
\]
This is a continuous, positive, symmetric curvature density of mass two. Thus each endpoint moment in \eqref{eq:fr-curvature} is one. It describes a fixed downstream decision envelope independent of the realized shared seed. Finite action sets approximate that envelope uniformly, as shown above. At each horizon choose an approximation error $o(n^{-1})$; the declared supremum permits any finite action count, so this sequence is admissible and remains independent of the realized seed.

For every seed-conditioned partition, \eqref{eq:fr-weighted} bounds its compression cost. Averaging the seed cannot reduce that bound. The acquisition term converges by \eqref{eq:fr-baselimit}. Therefore
\[
 \liminf nV^{\rm pub}_{n,K}
 \ge\int\psi(t)b(t)\,dt+
       \frac1{24\kappa^2}\left[\int\psi(t)^{1/3}\,dt\right]^3.
\]
The optimizing point density for this weighted problem is
\[
 \lambda(t)=\frac1{2\sqrt3\kappa\sqrt{C-t(1-t)}}.
\]
It is proportional to $\psi^{1/3}$. Pointwise,
\[
 b(t)+\frac1{24\kappa^2\lambda(t)^2}=C/2.
\]
Multiplying by $\psi$ and integrating shows that the lower bound equals $C$.

For the matching code, use a uniformly shifted lattice of spacing $1/(K-1)$ in the cumulative coordinate of $\lambda$, adding the two endpoints. For a posterior location away from an endpoint, the local squared distance to a random shifted boundary has mean $[12(K-1)^2\lambda(t)^2]^{-1}+o(K^{-2})$. Endpoint clipping can only reduce the leading nearest-boundary distance. Smoothness bounds the remaining curvature and conditional-mean correction by $O(K^{-3})$, uniformly. Equation~\eqref{eq:fr-cell} thus bounds the mean compression profile by $[24(K-1)^2\lambda(t)^2]^{-1}+o(K^{-2})$. Together with the acquisition term this is at most $C/(2n)+o(n^{-1})$. The mass-two curvature bound proves attainment. The discrete-posterior comparison again costs $o(n^{-1})$.

For $J$ equally spaced phases the leading mean-square nearest-boundary distance is at most $1+2/J^2$ times the continuous-phase value. The smooth remainder is uniform in phase. Taking any $J=J_n\to\infty$ proves the stated finite-seed attainment. The source-dependent message still has $K$ possible values conditional on the shared phase; the phase requires separate public storage.

\subsection{Exact decoding and finite precision}
For an actual retained probability interval $[l,r]$, the posterior mean is $[G_n(r)-G_n(l)]/(r-l)$, where, with $i=\lfloor(n+1)w\rfloor$,
\[
G_n(w)=\frac{i(i+1)}{2(n+1)(n+2)}+
\frac{i+1}{n+2}\left(w-\frac{i}{n+1}\right).
\]
At $w=1$ the last term is zero. This is the exact integral of the discrete step posterior, rather than the affine surrogate.

For the fixed positive finite-ratio regime, rounding each cell boundary by at most $\zeta_n$ changes the encoding on an event of probability at most $2K\zeta_n$. Taking $\zeta_n=o((nK)^{-1})$ makes the regret perturbation $o(n^{-1})$. Posterior means rounded to $o(n^{-1})$ change optimal conditional decision loss by at most twice that amount. The private uniform randomization inside each count interval can be replaced by a rational grid fine enough that its total interval-probability error is $o(n^{-1})$. Polynomially fine grids suffice when $K\asymp\sqrt n$. A growing finite phase alphabet attains the shared limit; its public storage is separate from the $K$ source-dependent message labels.

\subsection{The local allocation rule for regular posterior profiles}
\label{app:allocation-context}
This subsection derives the continuous-density allocation principle used in \cref{sec:resources}. The discrete-posterior result follows from the preceding proof. Let $Q_n$ have densities $f_n\to f$ in $C^1[0,1]$, uniformly positive and bounded, and suppose $n e_n^{\mathrm{acq}}\to a$ uniformly. Write an interval cell as $[u,v]$, its width as $\Delta$, and its posterior mean as $c$. Uniform $C^1$ bounds and positivity give $c=(u+v)/2+O(\Delta^2)$. The cell formula \eqref{eq:fr-cell}, applied to $f_n$, then yields
\[
 e_n^{\mathrm{comp}}(c)=\frac{f_n(c)\Delta^2}{8}+O(\Delta^3)
\]
uniformly over the cells. The same expression is the cell maximum, since the two branches meet at $c$, increasing to it and decreasing after it.

Let $\lambda\in C^1[0,1]$ be positive, bounded away from zero, and integrate to one. Partition its cumulative coordinate into $K$ equal intervals. Then $\Delta=[K\lambda(c)]^{-1}+O(K^{-2})$ uniformly. If $K/\sqrt n\to\kappa\in(0,\infty)$, the scaled total-profile peaks are
\[
 a(c)+\frac{f(c)}{8\kappa^2\lambda(c)^2}+o(1).
\]
Uniform continuity of $a$ and the cell bound extend the corresponding upper allowance to every threshold. By \eqref{eq:fr-envelope}, an affine allowance $C h_\rho(t)$ with $h_\rho(t)=\rho t+(1-\rho)(1-t)$ is achieved asymptotically whenever
\[
 a(t)+\frac{f(t)}{8\kappa^2\lambda(t)^2}\le C h_\rho(t)
 \quad\text{for every }t.
\]
Under a uniformly positive slack $C h_\rho-a$, equalizing this local allowance gives the normalized density proportional to $[f/(C h_\rho-a)]^{1/2}$. Its normalizing integral is chosen consistently with the cell budget; at a binding private-design constraint it equals $\sqrt8\,\kappa$. When acquisition is exact, $a=0$ and this reduces to the nonuniform supplied-posterior rule $\lambda\propto[f/h_\rho]^{1/2}$. For a merely continuous limit $a$, uniformly approximate the positive allocation density by smooth positive densities and allow an arbitrarily small additional regret allowance. The derivation establishes a local construction and its allocation rule. Sharp minimax constants additionally require global interval-code converses and control of endpoint cells; those conclusions are not invoked here.

\section{The collective Gaussian program}
\label{app:gaussian}
The posterior mean $Z=\E[\Theta\mid Y]$ is sufficient because $\Theta=Z+E$ and $E$ is independent of $Y$. Averaging an encoder of $Y$ conditionally on $Z$ preserves $(\Theta,U)$ and reduces its information cost to $I(Z;U)$.

For any encoder, put $K=\E\operatorname{Cov}(Z\mid U)$. Conditional-mean decoding is simultaneously optimal for every quadratic task, and its loss is $\tr[G(S+K)]$. The law of total covariance gives $0\preceq K\preceq A$. At finite information, $K$ is positive definite. Indeed, a null direction would be an exactly reconstructed nondegenerate Gaussian scalar, which has infinite mutual information with its exact value. Conditional maximum entropy and concavity of log determinant yield
\[
I(Z;U)=h(Z)-h(Z\mid U)
\ge \frac12\log_2\det(AK^{-1}).
\]
The inequality can also be obtained by adding a vanishing independent Gaussian perturbation before taking the entropy bound, covering singular conditional laws without a formal subtraction of infinite differential entropies.

Conversely, for $0\prec K\preceq A$, take independent $U'\sim N(0,A-K)$ and $N\sim N(0,K)$, and set $Z=U'+N$. Their joint law disintegrates to a Gaussian channel from $Z$ to $U'$, with conditional covariance $K$ and information exactly $\frac12\log_2\det(AK^{-1})$. This proves \eqref{eq:gaussian} for arbitrary, including nonlinear, channels.

For fixed finite $M$, the feasible set is convex because $-\log\det$ is convex and the Loewner cap is convex. It is compact: each eigenvalue is at most $\lambda_{\max}(A)$, while
\[
\lambda_{\min}(K)\ge
\frac{\det(A)2^{-2M}}{\lambda_{\max}(A)^{k-1}}>0.
\]
The task set may be replaced by its compact convex hull. The loss is bilinear and continuous, so finite-dimensional minimax gives a saddle pair $(K_*,\bar G_*)$ and
\[
\min_K\sup_G\tr G(S+K)
=\max_{\bar G\in\conv\mathcal G}\min_K\tr\bar G(S+K).
\]
Represent $\bar G_*$ as a convex combination of at most $k(k+1)/2+1$ tasks by Carath\'eodory's theorem. At the saddle point, its weighted value equals the maximal task value, so every positive-weight task in that representation is active at $K_*$. This proves the finite-certificate assertion without requiring uniqueness or an efficient search over $\mathcal G$.

In the isotropic experiment, Gaussian conditioning gives $S=bI$ and $A=aI$, with $b=\sigma^2/(m+\sigma^2)$ and $a=m/(m+\sigma^2)$. For scalar projection tasks, the loss is $b+\lambda_{\max}(K)$. The determinant constraint implies $\lambda_{\max}(K)\ge a2^{-2M/k}$, attained by that scalar times $I$. Equation \eqref{eq:gaussian-isotropic} follows.

\section{Nonlinear discovery and finite retention}
\label{app:nonlinear}
This section proves \cref{thm:nonlinear} by recovering an unknown representation from raw observations. All constants may depend on $\alpha,A,\sigma$, which remain fixed, but not on $d,r,m,M$.

\subsection{Exact metric and observation identities}
For jointly standard Gaussian $U,V$ of correlation $\rho$,
\[
\E\sin U\sin V
=\tfrac12\{e^{-(1-\rho)}-e^{-(1+\rho)}\}
=e^{-1}\sinh\rho.
\]
Also $\E\sin^2Z=e^{-1}\sinh1$ and, by integration by parts, $\E[Z\sin Z]=\E\cos Z=e^{-1/2}$. Normalization proves \eqref{eq:sine-kernel} and \eqref{eq:sine-moment}. Expanding the square proves \eqref{eq:sine-metric}. Since the conditional Gaussian responses have a common covariance, their one-sample divergence in nats is
\begin{equation}
D_{\rm KL}(P_{a,u}\Vert P_{b,v})
=\frac{1}{2\sigma^2}\norm{F_{a,u}-F_{b,v}}_2^2.
\label{eq:ridge-kl}
\end{equation}
This expectation includes the common Gaussian input distribution.

For each fixed projector $P$, choose an interior loading vector $a$ and perturb it by a sufficiently small nonzero vector in $\ker P$, keeping $u$ fixed. The two resulting representations have identical task response $PF$. Thus no fixed task identifies the family. For a fixed pair of representations, the output difference lies in $\operatorname{span}\{a,b\}$. Its second-moment matrix has rank at most two, and a rank-$r/2$ projector can include that range. Hence the largest projected population discrepancy for that fixed pair equals its full $L^2$ discrepancy. The converse evaluates two fixed complementary tasks on the same retained message before taking the training expectation.

\subsection{Regular geometry and covering}
On a small chart with $a$ near a loading of norm strictly between $\alpha$ and $A$, and $u$ near $e_1$, orient signs consistently. Set $\kappa(\rho)=\sinh\rho/\sinh1$. Then
\[
\norm{F_{a,u}-F_{b,v}}_2^2
=\norm{a-b}^2+2a^\top b\{1-\kappa(u^\top v)\}.
\]
On this chart $a^\top b$ is bounded above and below by positive constants and $\kappa'(1)=\coth1>0$. Since $1-u^\top v=\norm{u-v}^2/2$, the metric is uniformly equivalent to the product Euclidean metric. Parameterize $u(z)=(\sqrt{1-\norm z^2},z)$ on a fixed-radius ball and combine it with $r$ loading coordinates. This gives a $k=r+d-1$ dimensional chart whose constants do not grow with dimension.

Globally, the annulus of loadings has an $(C/\epsilon)^r$ cover and the unit sphere has an $(C/\epsilon)^{d-1}$ cover in Euclidean distance, with numerical bases independent of dimension. For example, the sphere bound follows from a maximal separated set and spherical-cap volume bounds; the excess factor polynomial in $d$ is absorbed into a numerical base to the power $d-1$. The response map is uniformly Lipschitz in the product coordinates because $g$ is Lipschitz and $X$ is isotropic. Thus the function class has a cover of size at most $(C/\epsilon)^k$. Quotienting the simultaneous sign change can only reduce that size.

\subsection{Reducing arbitrary decoders to one representation estimate}
Let a protocol have operational worst-task risk $R$. Choose two fixed complementary rank-$r/2$ projectors $P_0,Q_0$. Project each task decoder into its task range, which cannot increase loss. On the same retained message and public seed, evaluate both functions and add them. Orthogonality gives a full-function estimate $H$ with
\[
\E\norm{H-F_{a,u}}_2^2\le2R.
\]
If the decoders use private randomness, generate both tapes independently in this reduction; they carry no additional source information. Alternatively, average the tapes first by squared-loss convexity.

For a lower bound restricted to a compact local chart, choose a measurable $\eta$-nearest feasible function in that chart to $H$. Such a selection can be made from a fixed countable dense sequence, avoiding an unproved measurable exact-minimizer assertion. The true chart member is feasible, so the reconstructed function has error at most four times the error of $H$, plus a quantity tending to zero with $\eta$. Local metric equivalence therefore turns $R$ into a parameter mean-square error bound. The reduction uses two tasks only in the converse; the deployed protocol still faces whichever one task is announced independently.

\subsection{Acquisition and memory converses}
In the local $k$-chart, take a Euclidean packing of a ball of radius $b$ by points separated by a fixed fraction of $b$. Its logarithmic cardinality is at least $c_0k$. By \eqref{eq:ridge-kl}, every $m$-sample divergence to the center is at most $Cmb^2$. Choose
\[
b^2=c_1\left(1\wedge\frac{k}{m}\right)
\]
with $c_1$ sufficiently small. Fano's inequality then gives expected parameter error at least $cb^2$. The arbitrary-decoder reduction transfers this to task loss. The standard two-point form covers any bounded small dimension not covered by a chosen packing constant.

For retention, reveal the parameter for free. On a local chart let $Z$ be uniform on $[-a/\sqrt{k},a/\sqrt{k}]^k$ for a small fixed $a>0$. Any hard $M$-bit message $J$ with independent shared seed $R_0$ has
\[
I(Z;J,R_0)=I(Z;J\mid R_0)\le M.
\]
For its reconstructed chart coordinate $\widehat Z$, put $D=\E\norm{Z-\widehat Z}^2$. The entropy bound gives
\[
M\ge k\log_2(2a/\sqrt{k})
 -\frac{k}{2}\log_2\frac{2\pi eD}{k}.
\]
Thus $D\ge c2^{-2M/k}$, with $c$ dimension-independent. Source-independent decoder tapes do not change the mutual-information bound. The two-task reduction proves the memory converse. The maximum of the acquisition and retention lower bounds is at least half their sum.

\subsection{A complete raw-data estimator bound}
Let $B_*=\E[YX^\top]=a u^\top/\sqrt{\sinh1}$, and let $\widehat B=m^{-1}\sum_iY_iX_i^\top$. For unit vectors $s\in\R^r,t\in\R^d$, the variables $s^\top Y$ and $t^\top X$ are sub-Gaussian with uniform constants: the signal in the first is bounded by $A/\gamma$, and its noise is Gaussian. Their product need not be independent, but is sub-exponential by Cauchy--Schwarz for exponential moments. Scalar Bernstein gives
\[
\Prob\left\{\left|s^\top(\widehat B-B_*)t\right|
>C\left(\sqrt{x/m}+x/m\right)\right\}\le2e^{-x}.
\]
Take $1/4$-nets of the two unit spheres, with combined size at most $9^{r+d}$, and apply the rectangular operator-norm net comparison. A union bound gives
\[
\Prob\left\{\norm{\widehat B-B_*}_{\op}
>C\left(\sqrt{(r+d+x)/m}+(r+d+x)/m\right)\right\}\le2e^{-x}.
\]
Integrating the tail yields
\begin{equation}
\E\norm{\widehat B-B_*}_{\op}^2
\le C\left\{\frac{r+d}{m}+\left(\frac{r+d}{m}\right)^2\right\}.
\label{eq:crosscov-bound}
\end{equation}
The sole nonzero singular value of $B_*$ is at least $\alpha/\sqrt{\sinh1}$. Let $\widehat u$ be a leading right singular vector, and take the leading left singular vector times its singular value times $\sqrt{\sinh1}$ as the loading estimate; clip its norm into $[\alpha,A]$. If $\norm{\widehat B-B_*}_{\op}$ is less than one quarter of the signal singular value, elementary singular-vector perturbation gives product-coordinate error at most a fixed multiple of this squared operator error, up to the common sign. The metric identity transfers it to function error. Outside that event the clipped function loss is bounded by $4A^2$, and Markov's inequality bounds the exceptional probability. For $m\ge k$, \eqref{eq:crosscov-bound} is $O(k/m)$; for $m<k$, a fixed feasible estimate gives constant error. Consequently
\[
\sup_{a,u}\E\norm{F_{\widetilde a,\widetilde u}-F_{a,u}}_2^2
\le C\left(1\wedge\frac{k}{m}\right).
\]
Computing the sample cross-covariance costs $O(mrd)$ arithmetic operations. A dense singular-value decomposition costs $O(rd\min\{r,d\})$, absorbed by $O(mrd)$ in the informative regime $m\ge k$. Work here is real-arithmetic work; a finite-precision implementation must charge precision and approximation tolerances separately.

\subsection{Retention constructions and exact accounting}
An $\epsilon$-cover of the function class with at most $2^M$ centers exists at radius at most $C2^{-M/k}$, with a larger constant covering the small-$M$ regime. Quantize the learned function to a nearest center and answer the task by projecting that center's function. The squared triangle inequality and the estimator bound prove the upper half of \eqref{eq:ridge-law}. This construction provides an existential covering code; the explicit chart code below gives the stated polynomial-time alternative.

For a direct code, select an index $j$ of a largest-magnitude coordinate of $\widetilde u$ and orient that coordinate positively, flipping $\widetilde a$ simultaneously. The $d-1$ ratios $\widetilde u_i/\widetilde u_j$ lie in $[-1,1]$. Retain $j$, then quantize those ratios and the $r$ loading coordinates in $[-A,A]$ with $b$ bits per scalar, where
\[
b=\left\lfloor\frac{M-\lceil\log_2d\rceil}{k}\right\rfloor
\]
when $M\ge\lceil\log_2d\rceil$. Decode by normalization, and clip the loading into the ball of radius $A$. The number of source-dependent bits is at most $\lceil\log_2d\rceil+kb\le M$. Lipschitz normalization on vectors with norm at least one gives additional function error at most $Ck2^{-2b}$. For smaller $M$, retain a fixed function. This proves the declared exponent and makes the dimension-dependent overhead of the explicit code visible.

\subsection{The two orders of task selection}
For a positive semidefinite $r\times r$ matrix $C$,
\[
\frac qr\tr C\le\Psi_q(C)\le\tr C.
\]
The function $\Psi_q$ is convex, being a supremum of linear forms over rank-$q$ projectors. Thus
\[
\Psi_q(\E C)\le\E\Psi_q(C)\le\E\tr C
\le\frac rq\Psi_q(\E C),
\]
which proves \eqref{eq:quantifier-factor}. This calculation concerns a common function decoder $P\widehat F$. The lower-bound reduction above covers more general task-dependent decoders separately.

\section{Active-task and population certificates}
\label{app:stress}
Let $D_{\mathcal C},L_{\mathcal G}>0$ and set $\eta=D_{\mathcal C}/(L_{\mathcal G}\sqrt T)$. Projection onto the closed convex design set and expansion of squared distances give, for every $K\in\mathcal C$,
\[
\sum_{t=1}^T\langle G_t,K_t-K\rangle
\le\frac{D_{\mathcal C}^2}{2\eta}
+\frac\eta2\sum_t\norm{G_t}^2
\le D_{\mathcal C}L_{\mathcal G}\sqrt T.
\]
Add the $S$ term, minimize over the comparison $K$, and divide by $T$. Convexity of $f$ and the approximate maximization guarantee imply
\[
f(\bar K_T)-\min_K\langle\bar G_T,S+K\rangle
\le D_{\mathcal C}L_{\mathcal G}/\sqrt T+\overline\delta_T.
\]
Weak minimax duality gives the lower bracket, and feasibility gives the upper bracket in \eqref{eq:stress-gap}. Carath\'eodory applied to the returned matrices preserves $\bar G_T$ with at most $1+\operatorname{affdim}(\mathcal G)$ positive weights. The lower bound is unchanged. If $f(\bar K_T)$ cannot be evaluated exactly, one final $\delta$-accurate maximization call gives upper bound $\langle G, S+\bar K_T\rangle+\delta$ and adds $\delta$ to the certified gap. Degenerate cases $D_{\mathcal C}=0$ or $L_{\mathcal G}=0$ are immediate.

For projection tasks, Ky Fan's variational principle gives $f(K)=\Psi_q(S+K)$ and a top-$q$ eigenspace is a maximizer. The convex projection step is still an optimization subproblem. For the Gaussian feasible set, log determinant and the Loewner cap define it explicitly; no arbitrary nonlinear task-search procedure is presumed.

For validation, let $P_*$ maximize $\tr(P\Sigma)$ and $\widehat P$ maximize $\tr(P\widehat\Sigma)$ over rank-$q$ projectors, where $\Sigma=C+\sigma^2I$. The middle comparison below is nonpositive by empirical optimality:
\begin{align*}
\tr(P_*\Sigma)-\tr(\widehat P\Sigma)
&=\tr P_*(\Sigma-\widehat\Sigma)
 +\tr(P_*-\widehat P)\widehat\Sigma
 +\tr\widehat P(\widehat\Sigma-\Sigma)\\
&\le2q\norm{\widehat\Sigma-\Sigma}_{\op}.
\end{align*}
Subtracting the equal noise term $q\sigma^2$ proves \eqref{eq:validation}.

If $\sup_{\norm v=1}\norm{v^\top R}_{\psi_2}\le\tau$, each centered square $(v^\top R)^2-\E(v^\top R)^2$ has sub-exponential norm at most $C\tau^2$. Bernstein over a $1/4$-net of the output sphere yields, with probability at least $1-\delta$,
\[
\norm{\widehat\Sigma-\Sigma}_{\op}
\le C\tau^2\left[\sqrt{\frac{r+\log(2/\delta)}{N}}
+\frac{r+\log(2/\delta)}{N}\right].
\]
For bounded sine-ridge targets and candidates with bounded loadings, $\tau$ is bounded in terms of $A,\gamma,\sigma$. This applies conditionally on any candidate independent of the validation sample. The bound applies to a candidate independent of its validation sample; adaptive validation requires additional control.

\section{A computational obstruction with unchanged information}
\label{app:crypto}
For a length-preserving one-way permutation $f_n$, the two observations $X$ and $f_n(X)$ are bijective encodings of the same uniform source. Unrestricted kernels convert either into the other, so their Blackwell values agree. Both carry $n$ bits of mutual information about $X$.

An unrestricted encoder retains $X$ in $n$ bits and answers every parity. Conversely, exact service for the $n$ coordinate parities makes $X$ a function of the common retained message and public seed. With an independent seed $R_0$, $I(X;J,R_0)=n\le H(J\mid R_0)$, so the hard-message cost is at least $n$ bits. The accessible interface uses $O(n)$ work for encoding and for each requested parity.

Under the hidden interface, suppose polynomial-time encoder $E_n$ and decoder $D_n$ predicted an independent random parity with nonnegligible advantage. The composition
\[
A_n(y,r)=D_n(E_n(y),r)
\]
would be a probabilistic polynomial-time predictor of $\langle r,X\rangle\bmod2$ from $(f_n(X),r)$. This contradicts the Goldreich--Levin hard-core theorem \cite{goldreichlevin}. Randomized encoders, shared polynomial-time coins, and randomized decoders remain a probabilistic polynomial-time predictor. Retaining the full hidden observation is included. The theorem gives an asymptotic separation under the stated one-way permutation assumption and the specified cryptographic distribution.

\section{Verification and reproduction}
\label{app:verification}
General-purpose large language models were used for proof exploration, drafting, and adversarial auditing of the results in this manuscript.
The accompanying programs perform exact finite calculations, numerical evaluations of identities, and simulations of a specified estimator. The analytic proofs establish the arbitrary-encoder converses and infinite-family coding theorems.

\paragraph{Collective decision structure.}
Exact rational enumeration verifies the random-subset identity for 36 combinations of source size and subset size. One hundred stochastic finite channels verify the collision bound. Exhaustive pairwise comparisons at source dimensions $d=3,4,5,6$ verify that the coordinate and shared Boolean families have the same discrete task metric and fair individual responses. A direct finite-channel calculation checks the canonical reduction example: at crossover probability $0.17$, the raw and calibrated information costs differ by exactly one bit while their errors agree.

\paragraph{Binary coding.}
Numerical quadrature checks the defining integral for the private and shared constants at five positive ratios; the largest discrepancy is below $2\times10^{-11}$. A separate implementation evaluates 18 finite codes using the exact discrete posterior, with $n\in\{64,256,1024\}$ and three values of $K/\sqrt n$. The shared implementation uses 32 phases. The finite linear program optimizes the adversarial downstream decision for each specified code. These values report the risks achieved by the specified finite codes, alongside their separate asymptotic comparison.

\paragraph{Nonlinear discovery.}
Gauss--Hermite quadrature with 48 nodes per independent Gaussian variable verifies the sine correlation at nine correlations and the first-moment identity, with errors below $7\times10^{-16}$. Five hundred independently generated parameter pairs verify the fixed-pair collective projection identity; the maximum matrix discrepancy is below $3\times10^{-15}$. Five hundred covariance-mixture checks satisfy the factor-two order-of-operations inequality. Five hundred residual-covariance checks satisfy the deterministic empirical stress-task inequality; the maximum observed ratio of the task gap to its bound is below $0.13$.

The raw-data simulation uses $d=12$, $r=4$, noise standard deviation $0.5$, unit-norm true loadings, and the declared clipping interval $[0.5,1.5]$. At each of five sample sizes it generates 160 independent sources and training data sets. It compares the unquantized moment estimator with chart codes using one, three, or five bits per scalar, for total messages of 19, 49, and 79 bits. Population error is evaluated from the exact nonlinear kernel, rather than a fitted test curve. Every generated run is retained. The plotted intervals are marginal $1.96$-standard-error intervals across replications. The coarse code can be worse than a trivial response at small budgets; it is displayed without replacing unfavorable outcomes.

\paragraph{Symbolic checks.}
A separate Wolfram Language evaluation returns zero for the sine covariance identity, zero for the random-subset algebraic identity, $\coth1$ for the derivative of the normalized correlation kernel, and zero for the derivative of the binary integral primitive minus its integrand. The exact input and output are recorded. These checks establish algebraic identities only.

\paragraph{Nested-resolution checks.}
The new suite checks 240 stochastic encoders on binary and mixed-branch trees, with three independent seed values each, and ten explicit mask--prefix codes. Every task-risk lower bound and prefix upper bound passes to rounding precision. Five non-geometric profiles test entropy tails and crossing indices; further checks evaluate \eqref{eq:nested-loglog} and 24 finite dual-shattering constructions. Three of 960 rounded-reach cases have upward cost jumps, confirming why \eqref{eq:nested-profile} allows a comparable regular profile. The infinite-resolution theorem is established by the analytic argument in Appendix~\ref{app:nested}.

\paragraph{Decay-profile and parity checks.}
The decay-and-parity extension suite checks Parseval, the heavy-atom Fourier bound, and the conditional-entropy bound on 96 finite posterior distributions, including uniform, point-mass, concentrated-mixture, and fixed-seed Dirichlet cases. Twenty complete finite erasure channels verify every parity's error and the channel's information cost at the declared distortion. Sixteen finite block-decay cases check the entropy envelope; 5,001 plateau indices check the explicit exponent bounds, and six plateau endpoints check growing partial entropy sums. Twelve polynomial-profile crossing checks compare the prefix upper construction and single-scale Fano lower bound. These tests validate the listed finite formulas and implementations; the infinite-rate conclusions follow from the analytic proofs.

\paragraph{Symmetric parity-channel checks.}
The symmetric-channel suite enumerates 72 channels with $1\le k\le12$ and six distortion levels, including both endpoints. It checks the error of every nonzero parity, the entropy formula in \eqref{eq:parity-symmetric-finite}, and improvement over the explicit-erasure bound at interior distortion. For $k\le5$, direct joint-channel tables independently verify the mutual information and parity errors. Twenty additional evaluations test the fixed-interior-distortion expansion. These finite tests check the construction and its formulas; optimality follows from the analytic Fourier--entropy converse.

\paragraph{Exact parity converse.}
A separate suite tests the sharp entropy envelope on 3,840 fixed-seed posterior distributions across dimensions $k=1,\ldots,8$, including highly concentrated and nearly uniform laws. It also checks signed Walsh exponential moments and 49 complete symmetric-channel instances, including both distortion endpoints and $k=1$. An independent finalization suite exhausts small probability grids and all sign patterns at small dimensions, checks the entropy-dual optimizer and the interpolation nodes, and verifies equality for translated point-mass/uniform mixtures. The analytic majorant and entropy-duality argument establish the arbitrary-channel converse, with finite tests serving as numerical checks. The classical origin of the exponential-moment bound is recorded separately from the operational parity application.

\paragraph{Reproduction.}
The verification programs, numerical records, fixed random seeds, figure-generation code, and software requirements accompany this article as the ancillary file \texttt{reproducibility.zip}. After extracting that file, run the following commands from its root:
\begin{verbatim}
python code/verify.py
python code/verify_revision.py
python code/verify_nested.py
python code/verify_extensions.py
python code/verify_parity_refinement.py
python code/verify_exact_parity.py
python code/verify_finalization.py
python code/make_figures.py
bash build.sh
python code/check_release.py
\end{verbatim}
The ancillary archive contains the manuscript source, recorded outputs, dependencies, and file checksums. Floating-point replay is environment-dependent; the figures use LaTeX and vector graphics.

\end{document}